\documentclass[a4paper,11pt]{article}
\pdfoutput=1 

\usepackage{jheppub} 

\usepackage[T1]{fontenc} 
\usepackage{graphicx}
\usepackage{subfigure}
\usepackage{epsfig}
\usepackage{epstopdf}
\usepackage{color}
\usepackage{amsmath}
\usepackage[utf8]{inputenc}
\usepackage[mathscr]{euscript}
\usepackage{amsmath}
\usepackage{graphicx}
\usepackage{dcolumn}
\usepackage{bm}
\usepackage{epsfig}
\usepackage{amssymb,latexsym,mathrsfs}
\usepackage{graphicx}
\usepackage{color}
\usepackage{braket}
\usepackage{hyperref}
\usepackage{float}
\usepackage{diagbox}
\usepackage{textpos}
\newcommand{\sx}{\mathsf{x}}

\newcommand{\bo}{\raise-1mm\hbox{\Large$\Box$}}

\newcommand{\be}{\begin{equation}}
\newcommand{\ee}{\end{equation}}
\newcommand{\bea}{\begin{eqnarray}}
\newcommand{\eea}{\end{eqnarray}}

\title{Effects of Horizons on Entanglement Harvesting}

\author[a,b]{Wan Cong,}
\author[c]{Chen Qian,}
\author[d]{Michael R.R. Good,}
\author[a,b]{and Robert B. Mann}

\affiliation[a]{Department of Physics \& Astronomy, University of Waterloo,
\\Waterloo, Ontario, N2L 3G1, Canada}
\affiliation[b]{Institute for Quantum Computing, University of Waterloo,
\\Waterloo, Ontario, N2L 3G1, Canada}
\affiliation[c]{Department of Modern Physics,
University of Science and Technology of China,\\
Hefei, Anhui 230026, China}
\affiliation[d]{Physics Department \& Energetic Cosmos Laboratory, Nazarbayev University, \\Astana, 010000 Kazakhstan}

\emailAdd{wcong@uwaterloo.ca}
\emailAdd{qianch18@mail.ustc.edu.cn}
\emailAdd{michael.good@nu.edu.kz}
\emailAdd{rbmann@uwaterloo.ca}

\abstract{We study the effects of horizons on the entanglement harvested between two Unruh-DeWitt detectors via the use of moving mirrors with and without strict horizons. The entanglement reveals the sensitivity of the entanglement harvested to the global dynamics of the trajectories disclosing aspects of the effect that global information loss (where incoming massless scalar field modes from past null infinity cannot reach right future null infinity) has on local particle detectors. We also show that entanglement harvesting is insensitive to the sign of emitted radiation flux.}

\begin{document} 
\maketitle
\flushbottom

\section{Introduction}\label{intro}

Investigating the nature of horizons has a well-established history of revealing interesting physics. Perhaps the most prominent example is the Schwarzschild event horizon \cite{PhysRev.110.965} which helped confirm the existence of black holes. Studies using quantum field theory in curved spacetime revealed that the presence of horizons generically is accompanied by particle production; indeed particle creation is a fundamental phenomenon in curved spacetimes \cite{Birrell:1982ix,Parker:2009uva}, but can also take place in flat spacetimes in the frame of uniformly accelerated observers \cite{Davies:1974th,Unruh:1976db}, whose access to information is limited by their associated Rindler horizons. For a static black hole at late-times, the spectrum of the produced particles is that of a thermal blackbody \cite{Hawking:1974sw}. While resolving certain conundrums as to the role of the second law of thermodynamics, this phenomenon raises new paradoxes that have yet to be resolved \cite{Mann:2015luq}.
 
While it is known that particles are often created in entangled pairs, quantifying entanglement in quantum fields is not at all straightforward. An operational approach that is proving to be quite fruitful is entanglement harvesting \cite{Salton:2014jaa,Pozas-Kerstjens:2015gta}, which has its roots in the observation that atoms initialized as uncorrelated states can become entangled after some time due to the global nature of field correlators \cite{VALENTINI1991321}. The extraction of entanglement from the quantum vacuum has a number of interesting applications, including distinguishing a thermal bath from an expanding universe at the same temperature \cite{PhysRevD.79.044027}, probing the topology of spacetime \cite{Smith2016topology}, the discovery of separability islands (isolated regions of spacetime where harvesting is not possible) in anti-de Sitter spacetime \cite{Ng:2018drz,Henderson:2018lcy}, and the demonstration that black holes have `entanglement shadows' (a region about the black hole where entanglement extraction is extinguished) \cite{Henderson:2017yuv}.

Studying quantum entanglement in dynamical settings, for example during gravitational collapse, is considerably more difficult. 
A useful theoretical laboratory for studying such settings is that of \textit{mirror spacetimes}, pioneered in \cite{Davies:1976hi,Davies:1977yv}. The Dirichlet boundary condition imposed on the moving mirror in $(1+1)$ dimensions mimics the effect of gravity, but avoids the complications of extra dimensions, or curvature. This idealized setting allows one to more tractably  compute results and gain physical insight into the various phenomena \cite{fullingpage}. For example, certain limits of generic mirror trajectories can yield thermal responses \cite{Good:2013lca} as well as model a Schwarzschild black hole collapse from a null shell \cite{Good:2016oey,Good:2016atu}. More recently a study of entanglement harvesting from the vacuum of a massless scalar field in $(1+1)$ dimensions \cite{Cong:2018vqx} in moving mirror spacetimes indicated that entanglement shadows similar to those found for black holes \cite{Henderson:2017yuv} were present, and that the harvesting process was sensitive to the mirror trajectory, providing strong evidence that local detector measurements can distinguish between a collapsing black hole spacetime and an eternal black hole spacetime. Experimental observations  \cite{Johansson:2013asa} and recent proposals \cite{Chen:2017prl,Chen:2020sir} of the dynamical Casimir effect (DCE) further motivate the use of mirror spacetimes \cite{Good:2020byh,Good:2020uff,Good:2020fjz} to study aspects of particle creation in quantum field theory. 
  
The most popular mirror model studied in the literature is an initially inertial mirror that starts to accelerate to the left at $t=0$ and becomes asymptotically null, with $dx/dt\rightarrow -1$ as $t\rightarrow \infty$. The appeal of this mirror lies in its late time exactly thermal radiation, which arises thanks to the presence of a horizon. However, such horizon mirrors demonstrate numerous pathologies, such as infinite particle count, infinite energy and divergent entropy flux (see e.g. \cite{Good:2018aer,Fabbri, Good:2019tnf,Good:2018zmx}).


In a succinct fashion, we ask here the question: \\

{\it{Does the presence of a horizon substantially affect entanglement harvesting?}}\\

To do this, we study a family of mirror trajectories, parameterised by $\xi\in [0,1]$, the asymptotic speed of the mirror. A horizon is present only when $\xi=1$. While the existence of an horizon can be inferred from the entanglement harvested from the detectors, we find that the process is
 somewhat subtle. For all $\xi<1$ mirrors, the amount of entanglement between two detectors switched on for a finite time interval at some coordinate time $T$ always asymptotes to finite values at large $T$. In contrast, that between two detectors when $\xi=1$ does not; we show numerically that for a certain parameter choice, it increases linearly with time at large $T$. 
 
 As a bonus, since horizonless mirrors must emit negative energy flux (as we prove via Eq.~(\ref{sum}) in Sec.~\ref{nef}), and the horizon-possessing Schwarzshild mirror \cite{Good:2016atu} does not emit negative radiation; distinguishing between horizon and horizonless trajectories via harvesting can tell us about whether or not the associated entanglement measure can act as a probe into the nature of negative energy flux (NEF) \cite{Ford:1990id}.

The outline of our paper is as follows:  We introduce the set-up of entanglement harvesting with Unruh-DeWitt (UDW) detectors in Sec.~\ref{harvest}, and a class of horizonless mirrors that correspond to black hole collapse but evolve at ultra-late times to remnant states in Sec.~\ref{BHCdrift}.  We then discuss the results in Sec.~\ref{results}, with an emphasis on the effect of the horizon on concurrence with respect to the death zone in Sec.~\ref{effect}.  In Sec.~\ref{nef}, we prove NEF must be emitted by all horizonless mirrors and demonstrate the insensitivity of concurrence to the sign. In Sec.~\ref{conc}, we conclude.  Units are $\hbar = c = 1$. 

\subsection{Entanglement Harvesting with UDW detectors}\label{harvest}

The Unruh-DeWitt (UDW) detector model describes the interaction of a two level quantum system (the detector) with the quantum field. In this paper, we are interested in studying two identical detectors, which will be labelled as $j = A,B$. To describe the interaction between the detectors and the field, let $\hat\mu_j(\tau) = e^{i\Omega \tau}\hat\sigma_j^++e^{-i\Omega \tau}\hat\sigma_j^-$ denote the monopole moments of the detectors, with $\sigma_j^{+} = |e\rangle \langle g|,\sigma_j^{-} = |g\rangle \langle e|$ being the ladder operators, $\Omega$ the energy gap of the detectors and $\tau$ the proper time of the detectors. Since the space is flat and the detectors are both inertial, they have identical proper time.
In terms of these, the interaction Hamiltonian in the interaction picture is given by
    \begin{align}
    \label{eq: int}
        H_I^j(\tau) = \lambda\chi_j(\tau)\hat\mu_j(\tau)\otimes\hat\phi(\sx_j(\tau))\,, \hspace{0.5cm} j = A,B\,.
    \end{align}
    Here, $\chi_j(\tau)$ is a switching function that controls the strength of the interaction over time, $\hat{\phi}(\sx_j(\tau))$ is the field operator evaluated along the trajectories $\sx_j(\tau)$ of the detectors and $\lambda$ is the detector-field coupling strength. 

If we initiate the detectors in their ground states and the field in the vacuum state then at the end of the interaction via~\eqref{eq: int}, there is a non-zero probability of finding the detectors in their excited states. This probability depends on the background spacetime and can be used to extract non-local information about the the spacetime \cite{PhysRevD.94.104041, harvest, Smith_2014,Smith2016topology,Birrell:1982ix}. Furthermore, the two detectors can become entangled at the end of the interaction even though there was no direct interaction between them. The detectors are said to have \textit{harvested entanglement} from the field.

Since the detectors are 2-level systems, we shall employ the concurrence, an entanglement measure  defined for qubits that determines the entanglement of formation  \cite{EoF}. Since  entanglement of formation is a monotonically increasing function of concurrence, it is sufficient to compute the concurrence $\mathcal{C}$ of the end state of the detectors to quantify the  amount of entanglement between them.

This can be computed using standard perturbation theory, perturbing in $\lambda$, and the result is \cite{Smith2016topology}
    \begin{equation}
        \mathcal{C}(\rho_{AB}) =2\max\left\{0,|X|-\sqrt{P_AP_B}\right\}+O(\lambda^4) \,.
        \end{equation}
  In this expression, $P_j$ is the excitation probability of the detectors and $X$ is a measure of  their (non-local) correlation. To leading order in $\lambda$, they are
    \begin{align}
        X &= -\lambda^2\iint d t\,d t' \chi_A(t)\chi_B(t') e^{-i\Omega(t+t')} \bigg[\Theta(t'-t)W(\sx_A(t),\sx_B(t'))+\Theta(t-t')W(\sx_B(t'),\sx_A(t))\bigg]\,,\label{eq: x} \\
        P &= \lambda^2\iint  d t\,d t' \chi(t)\chi(t') e^{-i\Omega(t-t')}W(\sx(t),\sx(t'))\,,\label{eq: p} 
    \end{align}
    where $W(\sx,\sx')=\braket{0|\hat\phi(\sx)\hat\phi(\sx')|0}$ is the pullback of the Wightman function to the detector trajectories and $\Theta(\cdot)$ is the Heaviside step function. In this paper, we will be using a compact switching $\chi(\tau)$,
    \begin{equation}
    \label{eq: switch}
        \chi(\tau) = \begin{cases} \cos^4(\eta(\tau-T))\,, & -\frac{\pi}{2\eta}< \tau-T<\frac{\pi}{2\eta} \\
        0\,, & \text{otherwise},
        \end{cases}
    \end{equation}
which   peaks at $\tau=T$ and is zero outside the interval $[T-\frac{\pi}{2\eta},T+\frac{\pi}{2\eta}]$. It has a shape similar to a Gaussian switching function  used in  previous investigations \cite{Cong:2020crf,Ng:2016hzn}, and ensures   that the detectors   cannot be causally influenced by the late-time state of the mirror. The expressions in Eqs.\eqref{eq: p} and \eqref{eq: x} must be computed numerically. The numerical scheme used as well as comments on the numerical precision can be found in Appendix \ref{numerical convergence}.

\subsection{Asymptotically drifting mirrors}\label{BHCdrift}
\begin{figure}
\centering 
\includegraphics[width=3.2in]{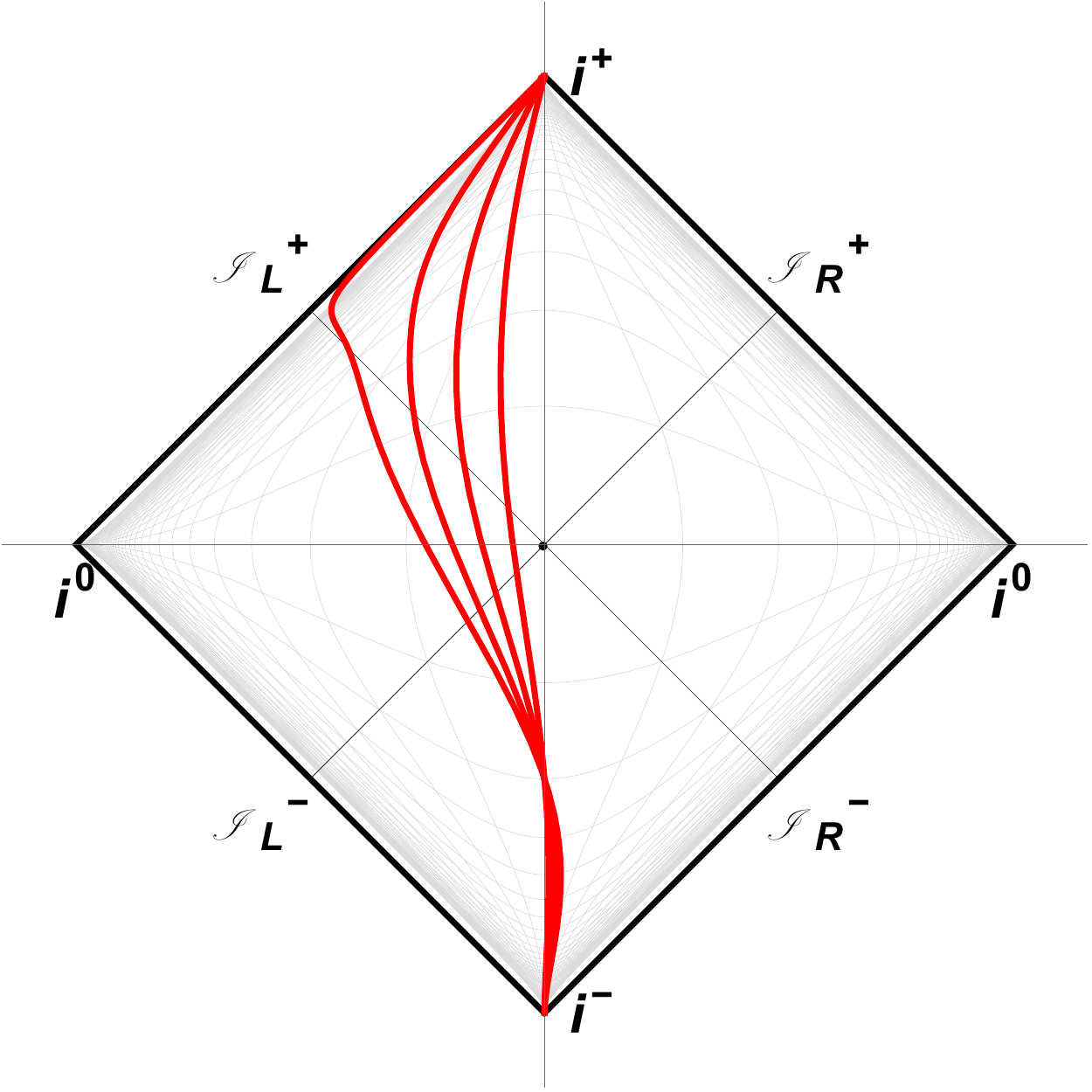} 
\caption{The trajectory, Eq.~(\ref{eq:traj}), plotted in a Penrose conformal diagram. Here $\kappa = 1$ and $\xi = 0.25$, $0.50$, $0.75$ and $0.99$ respectively. The horizon location $v_H = 0$. Notice that the mirror never forms a strict horizon as long as $\xi<1$: all left-moving modes ultimately reflect and become right-movers. 
} 
\label{fig:Penrose_Plot} 
\end{figure}

The moving mirror \cite{DeWitt:1975ys,Davies:1976hi,Davies:1977yv,Birrell:1982ix} is an accelerated boundary in flat spacetime that perfectly reflects field modes, creating particles that carry energy, with a similar production mechanism as that of light from black holes.  The mirror itself is a Dirichlet boundary condition imposed on the field equation of motion.  Often, the trajectory of the mirror is expressed in null coordinates, $u=t-x$ and $v=t+x$, due to the simplification associated with incorporating the dynamics into the field modes and their Doppler shift.  The mirror trajectory we will be working with is $v=p(u)$, with
\begin{equation}
\label{eq:traj}
    p(u)=u+\frac{\xi}{\kappa}\log\bigg[\frac{1+\xi}{2}W\big(\frac{2}{1+\xi}e^{\frac{2\kappa(v_H-u)}{1+\xi}}\big)\bigg]\,,
\end{equation}
where $W(\cdot)$ is the product-log function. The parameter $\xi$ represents the asymptotic final future speed of the mirror, see e.g. \cite{Good:2018ell}, while $\kappa$ parametrizes how fast this speed is achieved (it sets the scale of the system). The last parameter $v_H$ simply translates the mirror trajectory in time along the $t$ axis. This is seen more easily by writing the mirror trajectory as
\begin{equation}
    x_m(t) = \xi(v_H-t)-\frac{\xi}{2\kappa}W\big(2e^{2\kappa(v_H-t)}\big)\,.
\end{equation}
When $\xi<1$, the mirror drifts at constant velocity in the far future; 
the trajectory is an asymptotic inertial  version of the asymptotically null ``black hole collapse trajectory'' (BHC), which is obtained by setting\footnote{An eternal drifting light speed boundary produces zero particles as derived in Appendix A.} $\xi =1$ in Eq.~(\ref{eq:traj}),
\be \label{bhc} p(u)_{\textrm{BHC}} = v_H - \frac{1}{\kappa}W(e^{-\kappa(u-v_H)}). \ee
This trajectory has a one-to-one correspondence with the canonical case of time dependent particle creation from a collapsing star (null shell) \cite{Good:2016oey}. When $\xi=1$, $v=v_H$ corresponds to the location of the horizon of the mirror. This is the line beyond which left moving wave modes will reach $\mathcal{I}^+_{L}$ instead of $\mathcal{I}^+_{R}$.  The presence of a strict horizon signals information loss, as an observer on the right will never see information about the field modes that never get reflected.  These field modes are analogous to those that get trapped in a black hole, reaching the singularity, never to return again.  Left moving modes that do reflect are analogous to those waves that flow through the center of the collapsing star to eventually escape, reaching an outside observer.  In the extreme horizon case, an eternally drifting mirror at the speed of light proves to belong to the trivial case of inertial eternal constant velocity trajectories that do not radiate (see Appendix \ref{light}).

\section{Results}\label{results}

\subsection{Effect of horizon on concurrence}\label{effect}
 The BHC mirror has been shown  to give rise to an entanglement shadow (or `death zone' \cite{harvest}), similar to what was observed outside the $(2+1)D$ BTZ black hole \cite{Henderson:2017yuv}. While this seemed like an interesting correspondence between two scenarios when horizons are present, we find that in the mirror case, the death zone is not directly due to the presence of an horizon -- it is present even for mirrors with $\xi<1$.
 
 \begin{figure}
    \centering
    \includegraphics[scale=0.35]{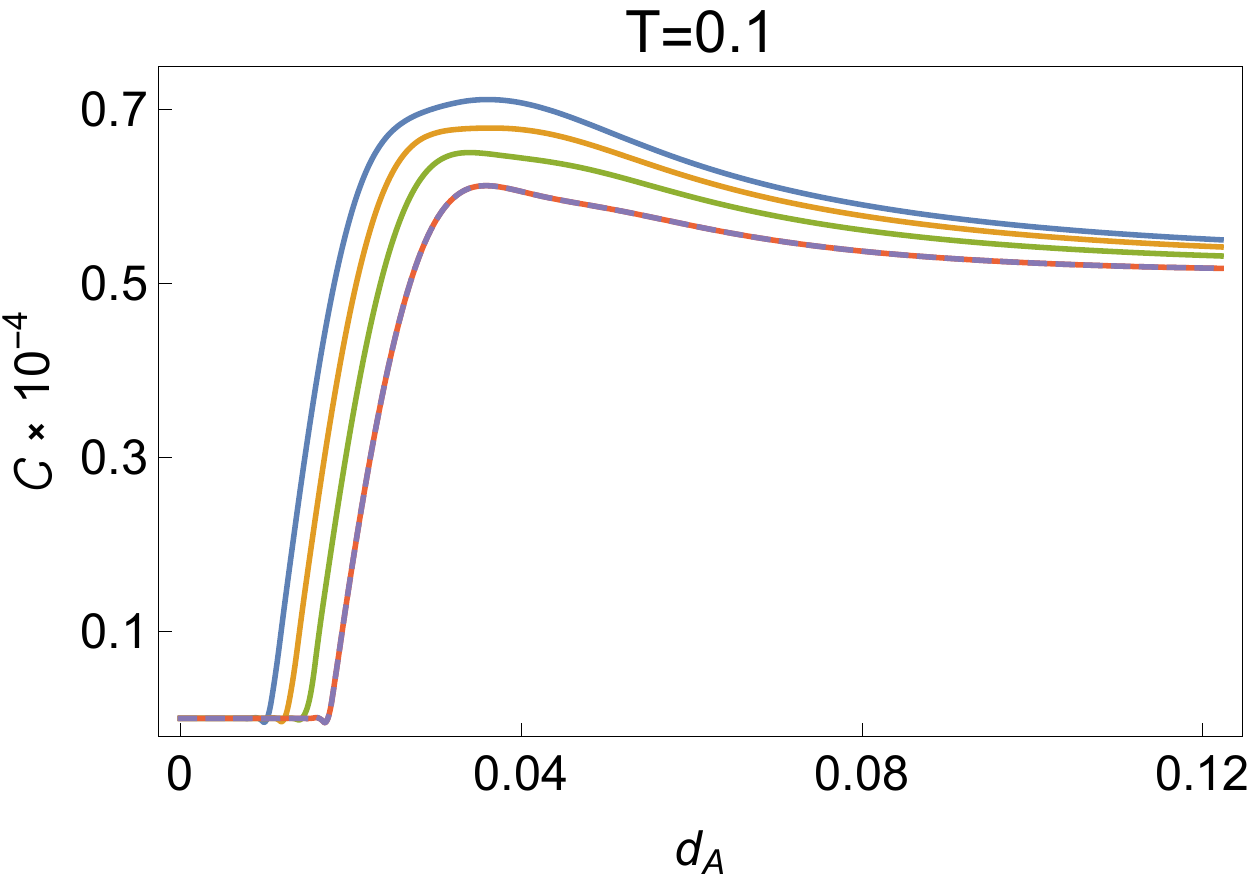}
    \includegraphics[scale=0.35]{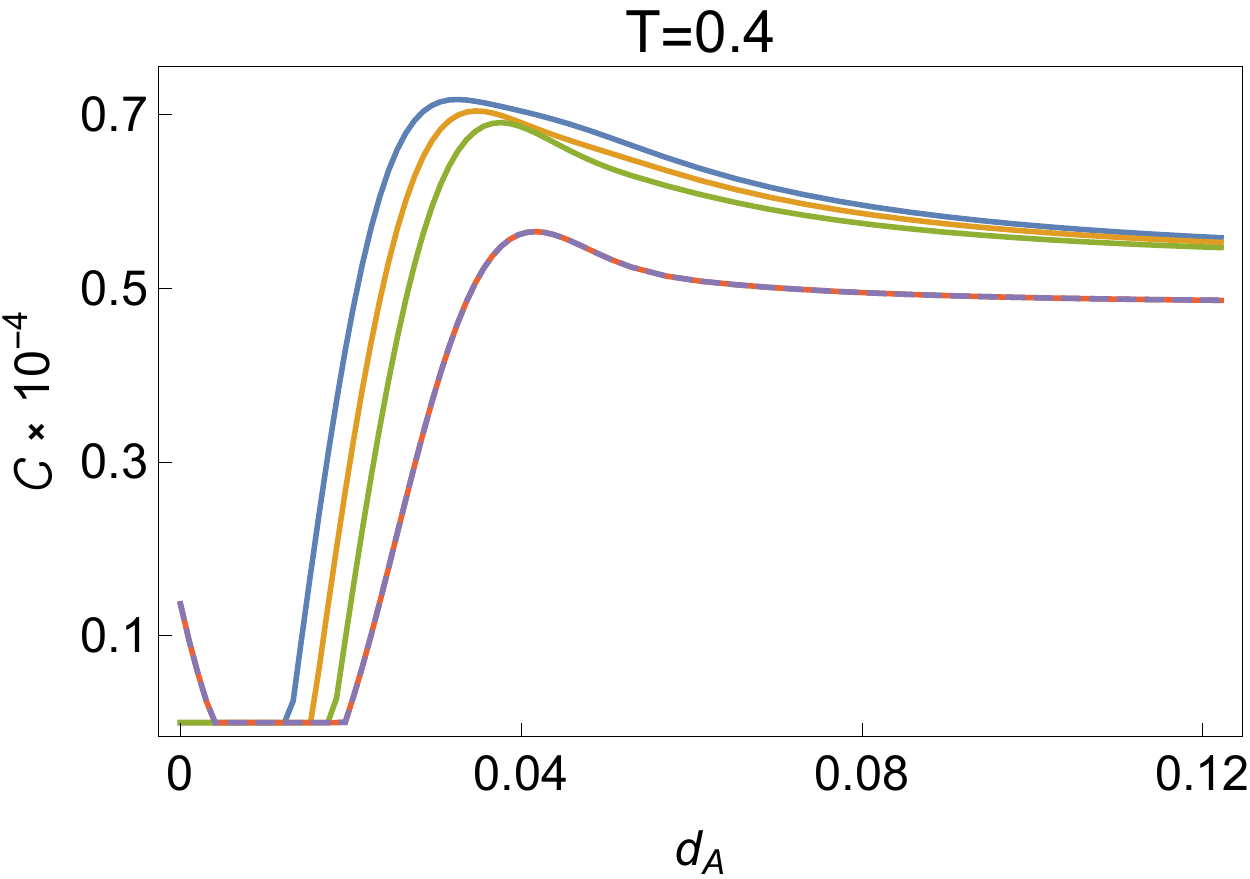}
    \includegraphics[scale=0.35]{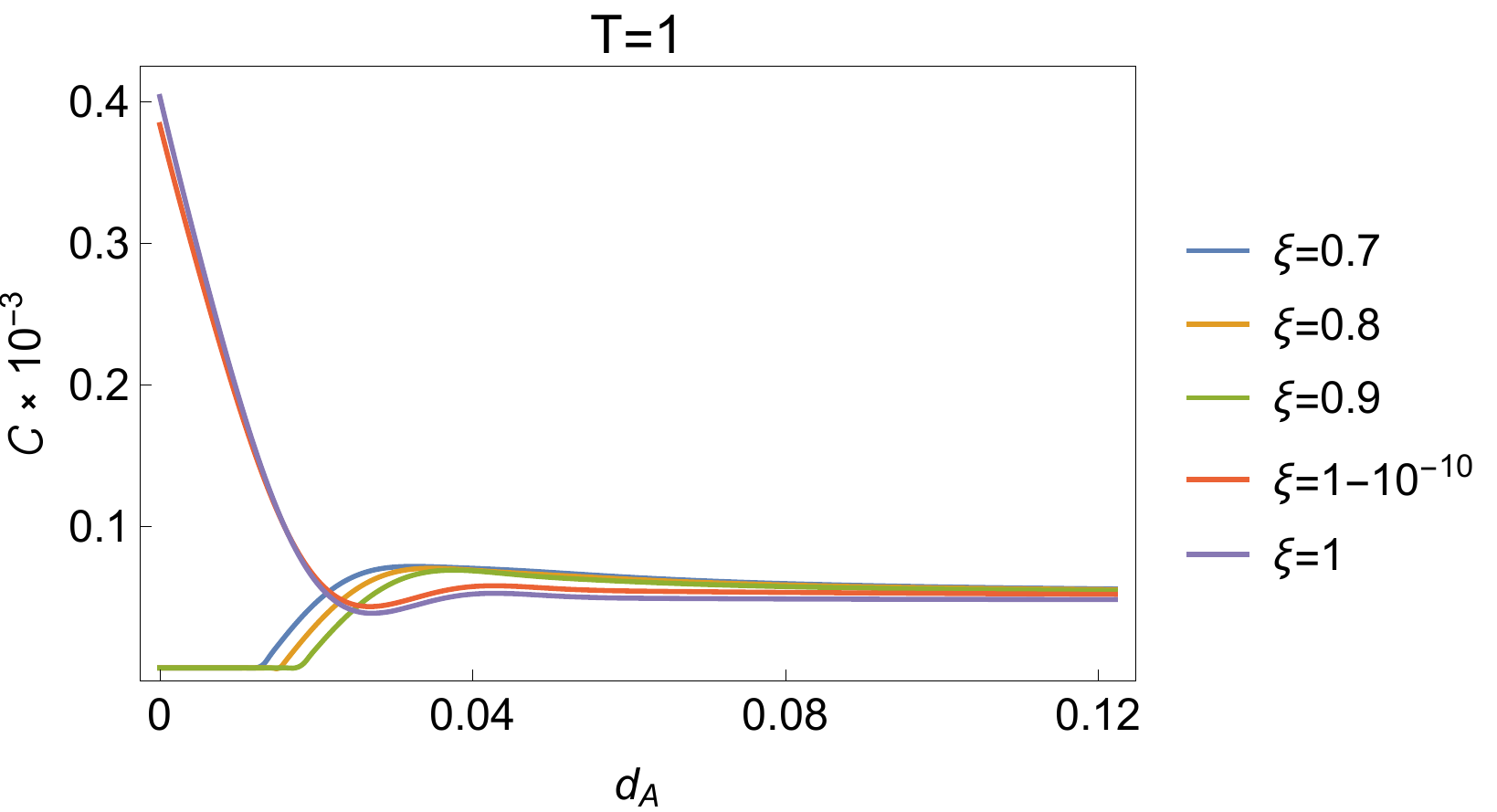} 
    \vspace{0.5cm}
    \includegraphics[scale=0.35]{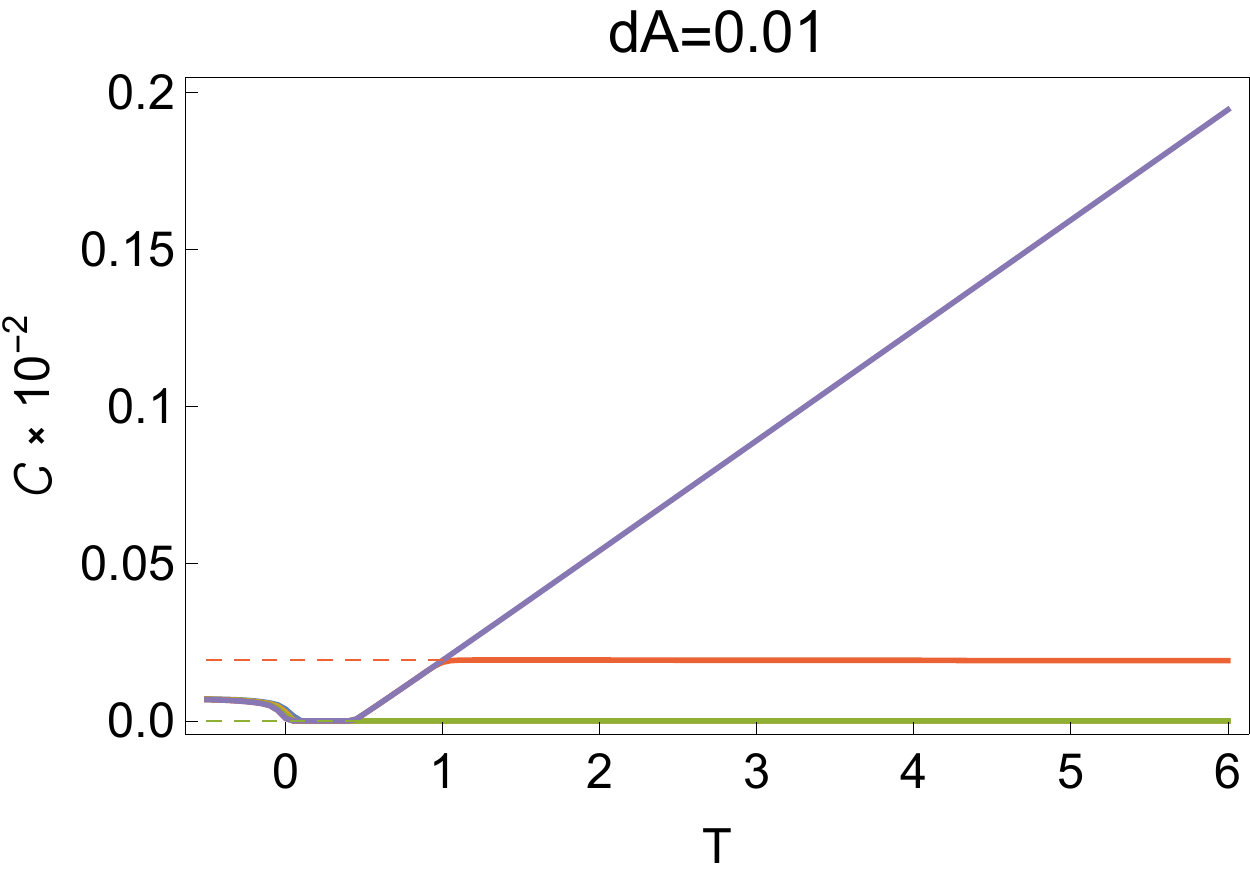}
    \includegraphics[scale=0.35]{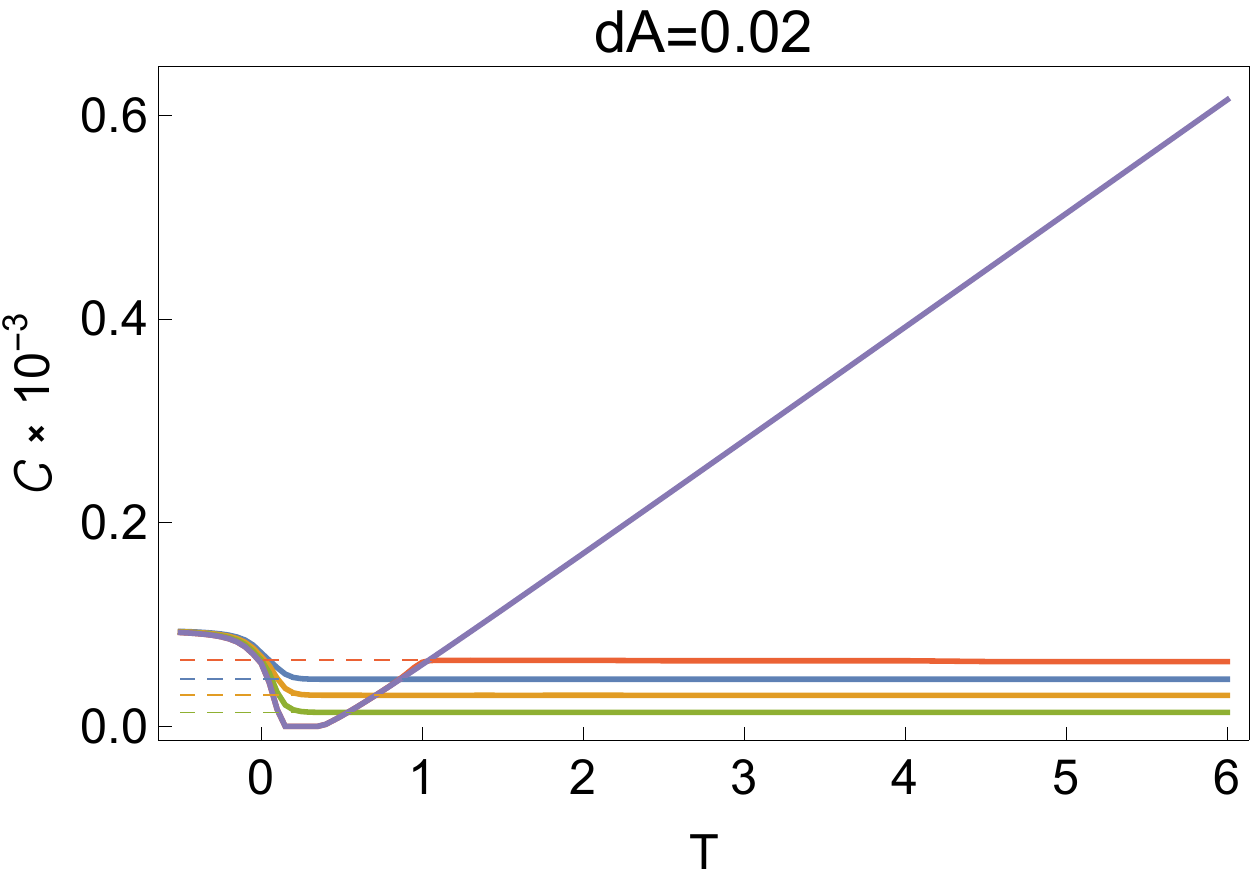}
    \includegraphics[scale=0.35]{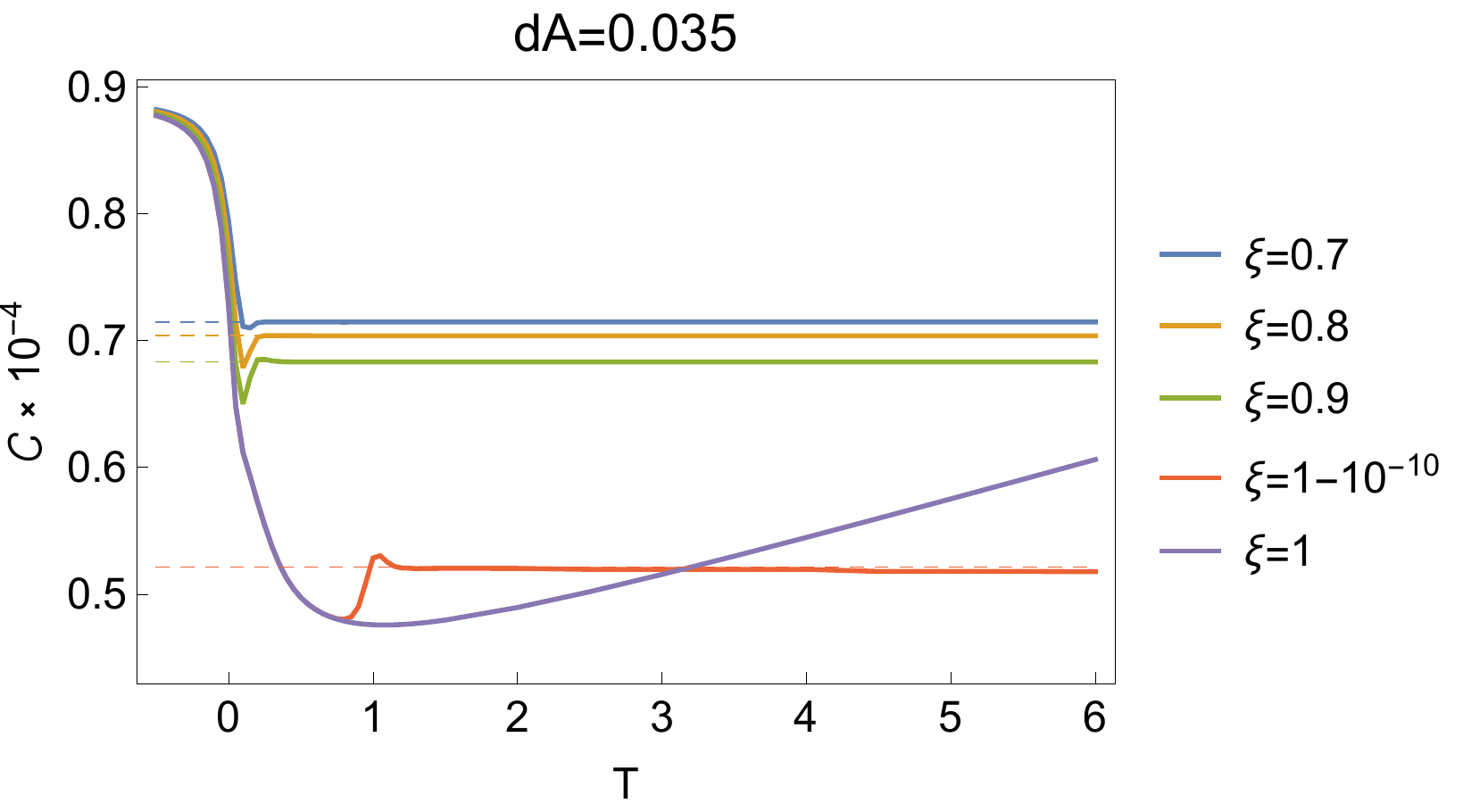}
    \vspace{0.5cm}
    \includegraphics[scale=0.35]{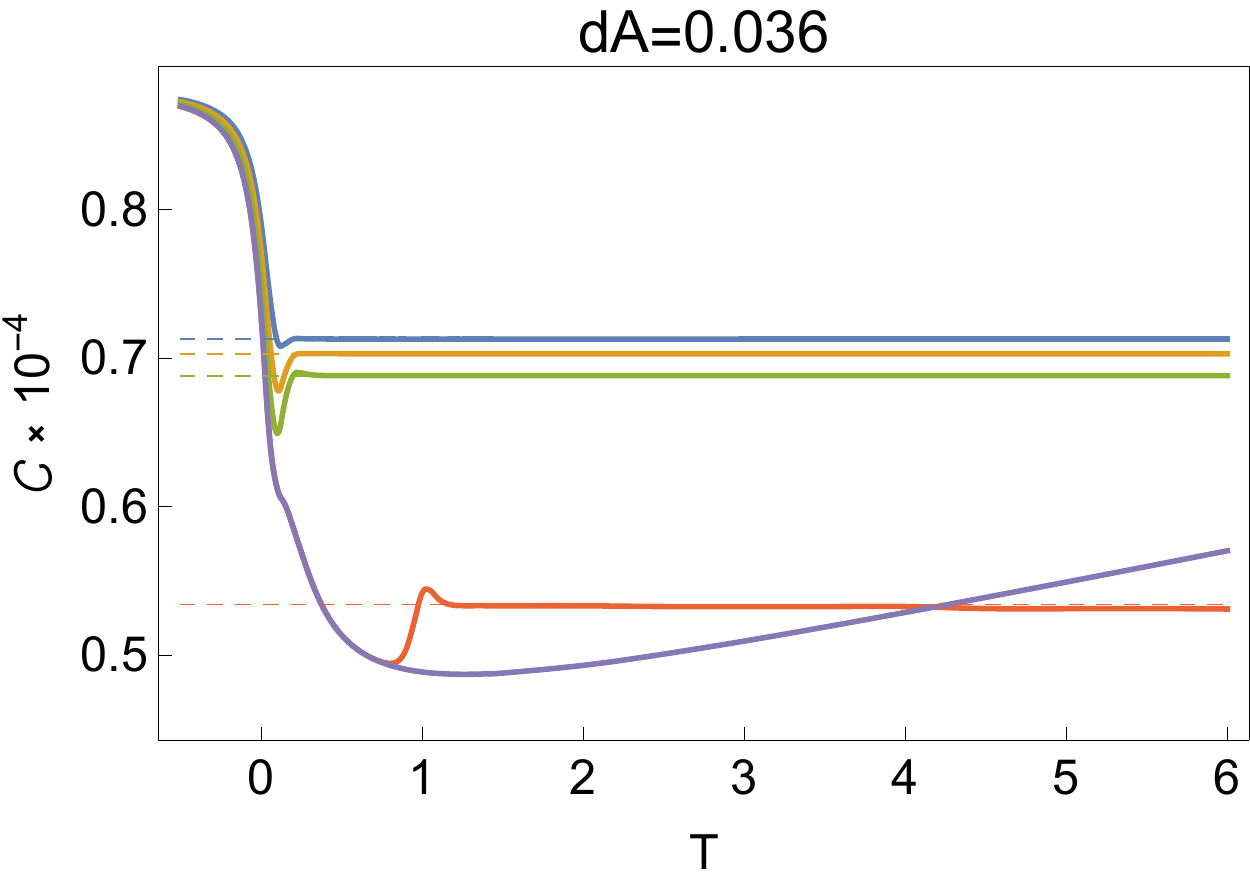}
    \includegraphics[scale=0.35]{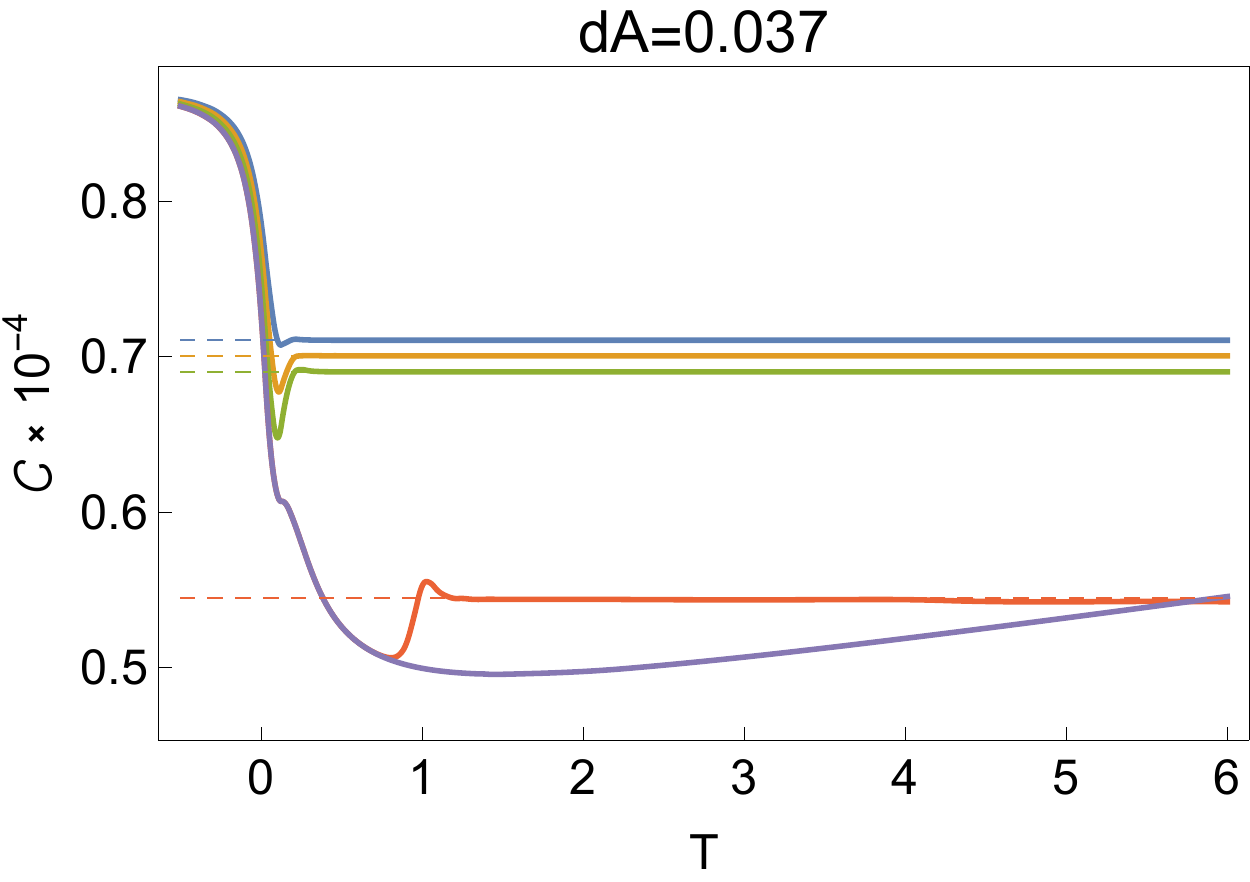}
    \includegraphics[scale=0.35]{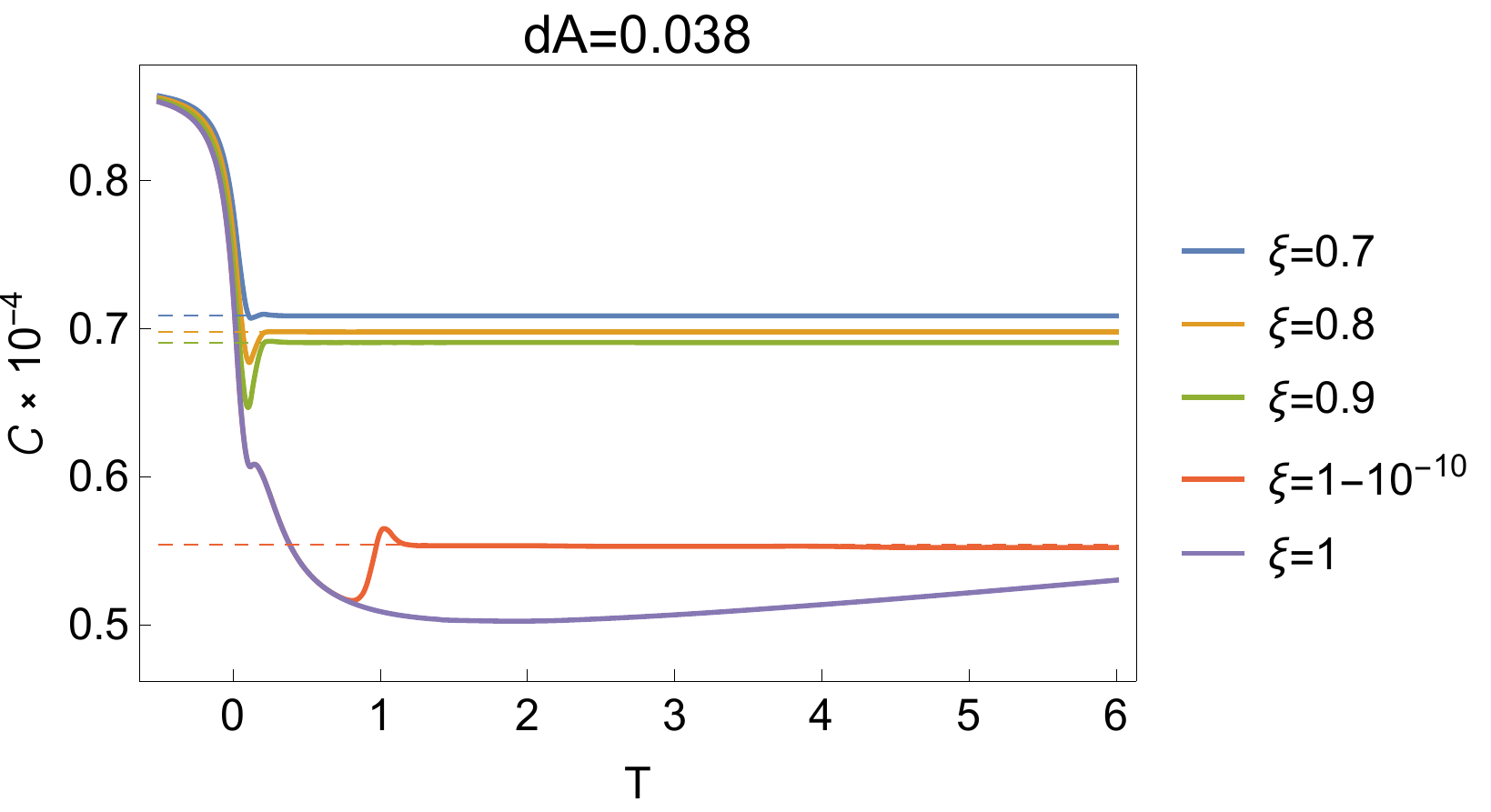}
    \caption{Effect of the parameter $\xi$. The parameter $\xi$ represents the asymptotic speed of the mirror moving to the left at late times. For $\xi = 1$, the mirror is asymptotically null and an ``horizon" appears at $v=0$. In the figures here, we investigate the effect of increasing $\xi$ on the concurrence between two UDW detectors. \textbf{Top:} Plots of concurrence against $d_A$ at times $T = 0.1,0.4,1$. The results at $T=0.1$ are similar to that obtained previously in \cite{harvest}, namely at small $d_A$, there is an entanglement death zone with zero concurrence, while the concurrence increases to some peak before asymptoting to some constant value at large $d_A$. However, we note that the death zone is present even for the $\xi<1$ mirrors, indicating that it is not a feature unique to mirrors with horizons. At a later time $T = 0.4$, we see a revival of the concurrence near the mirror when $\xi$ is close to $1$. Indeed, at $T=1$, we see that the death zone disappears completely for the $\xi=1$ mirror. \textbf{Middle/Bottom:} Plots of concurrence against $T$ for various $d_A$. Note that the curves for $\xi=0.7,0.8,0.9$ on the middle leftmost plot overlaps on the $T$-axis, corresponding to the fact that $d_A = 0.01$ lies within the entanglement death zone for these mirrors.
    The parameters used here are $\kappa = \sqrt{48\pi}$, $\eta = 23$, $\Omega = 50$ and $\Delta x = 0.05$. 
  }
    \label{fig: ccrvd}
\end{figure}

Let us be more specific. Consider the scenario in which detector $A$ (respectively $B$), placed at a fixed $x = x_A$ ($x_B$) to the right of the mirror, is switched on with the switching~\eqref{eq: switch}, peaking at some time $t=T$. As explained in the previous section, the two detectors can become entangled at the end of the interaction. In \cite{harvest}, it was found that as $d_A = x_A-x_m(T)$ decreases (i.e., as detector $A$ gets closer to the mirror) while keeping the detector separation $\Delta x = x_B-x_A$ fixed, concurrence will decrease until it reaches $0$ at some critical $d_A$. In other words, for a given mirror and fixed $\eta$, $T$, $\Delta x$ and $\Omega$, there is a minimal $d_A$ below which it is impossible to entangle the detectors. This region is the entanglement death zone. The purple curve in the top left plot of Fig. \ref{fig: ccrvd} illustrates this for $\xi=1$. However, the death zone is not unique to this BHC mirror. For example, when $\xi=0.7$ (blue), a death zone also exists. In fact we have checked that such entanglement death zones are present even when the mirror is moving with constant, non-zero velocity, and therefore may more generally be a characteristic of non-static mirrors. 

Our results are commensurate with previous studies on entanglement death.
While the presence of horizons certainly plays a role
\cite{Henderson:2017yuv}, other factors are also at play, including the state of motion of the detectors
\cite{Lin:2008jj,Ostapchuk:2011ud}, boundary conditions \cite{Cong:2018vqx,Cong:2020crf}
local vs. global considerations
\cite{Martin-Martinez:2015qwa}, and sensitivity of the detector to particular parameterizations
\cite{Ng:2018drz,Henderson:2018lcy}.   In particular, higher dimensional effects on entanglement death will be an interesting question to address, since the evidence we present here is limited to (1+1)-dimensions.

In addition, we observe a ``revival'' of entanglement close to the mirror at later times. This is illustrated in the $T=0.4$ and $T=1$ plots on the first row of Fig. \ref{fig: ccrvd}: at $T=0.4$, there still exists regions of entanglement death for all $\xi$ values, but at $T=1$, the death zone disappears for near-null mirrors. At this $T$, we have checked numerically that entanglement harvesting at $d_A=0$ becomes possible, i.e. $\mathcal{C}>0$, when $\xi\gtrsim 0.9997$.


From plotting $\mathcal{C}$ as a function of $d_A$, we saw that a revival in entanglement close to the mirror is possible at intermediate times while the death zone can disappear completely at late times. However, once again this is not unique to the BHC mirror, since the same features are observed for the $\xi=1-10^{-10}$ mirror. 

A distinct difference in behaviour arises when we consider a second scenario: once again, the detector separation $\Delta x$ is fixed but instead of varying $d_A$, we fix this and consider the effect of varying $T$. Some representative cases are shown in the middle and bottom plots of Fig. \ref{fig: ccrvd}. From these plots we see that while the concurrences for $\xi<1$ mirrors asymptote to finite values at large $T$, an asymptote does not seem to be present for the BHC mirror. In all cases, the $\mathcal{C}$ in the BHC mirror spacetime seem to increase linearly at late times\footnote{Due to computational constraint, we checked this up to $T\approx30$, where the UV regulator $\epsilon$ in the Wightman function (see Eqn.~\eqref{eq:wightman} in Appendix) needs to be $\sim10^{-320}$ for convergence.}. This is clearly visible in the middle row of  Fig.~\ref{fig: ccrvd}; it is also present in the bottom row, though a longer plotting range in $T$ is required to more dramatically see the effect.

At late times, the mirrors approximately move at constant speeds. It is thus natural to expect the large $T$ asymptote for the $\xi<1$ cases to be equal to the concurrence of detectors situated in spacetimes with mirrors moving (eternally) at the corresponding constant speeds. In fact, the dashed lines representing the asymptotic values of the different mirrors in Fig. \ref{fig: ccrvd} are obtained precisely in this way. We also show how this result can be obtained analytically in Appendix \ref{asymptotic}. Roughly speaking, when $\xi<1$, both the $P$ and $X$ terms have finite large $T$ limits, corresponding to the constant speed mirror $P$ and $X$ values. This thus gives an asymptotic value for the concurrence. However corresponding values for $\xi=1$, do not exist. As we show in the appendix, both $P$ and $X$ $\rightarrow \infty$ at large $T$ when $\xi=1$. We note that the results obtained are only valid to leading order in perturbation. In particular, the apparent linear increase in $P$, $|X|$ and $\mathcal{C}$ in $T$ for the $\xi=1$ mirror at late times will not continue indefinitely in the real world --- perturbations of higher orders will eventually be needed to accurately describe the large $T$ behaviour.

We can obtain some physical intuition for this result by noticing that the last rays of light that hit the mirror are extremely red-shifted to observers at infinity.   Since the mirror is receding and the wavelengths get stretched out, the linear growth in concurrence is almost certainty related. 
 
\begin{figure}
    \centering
    \includegraphics[scale=0.5]{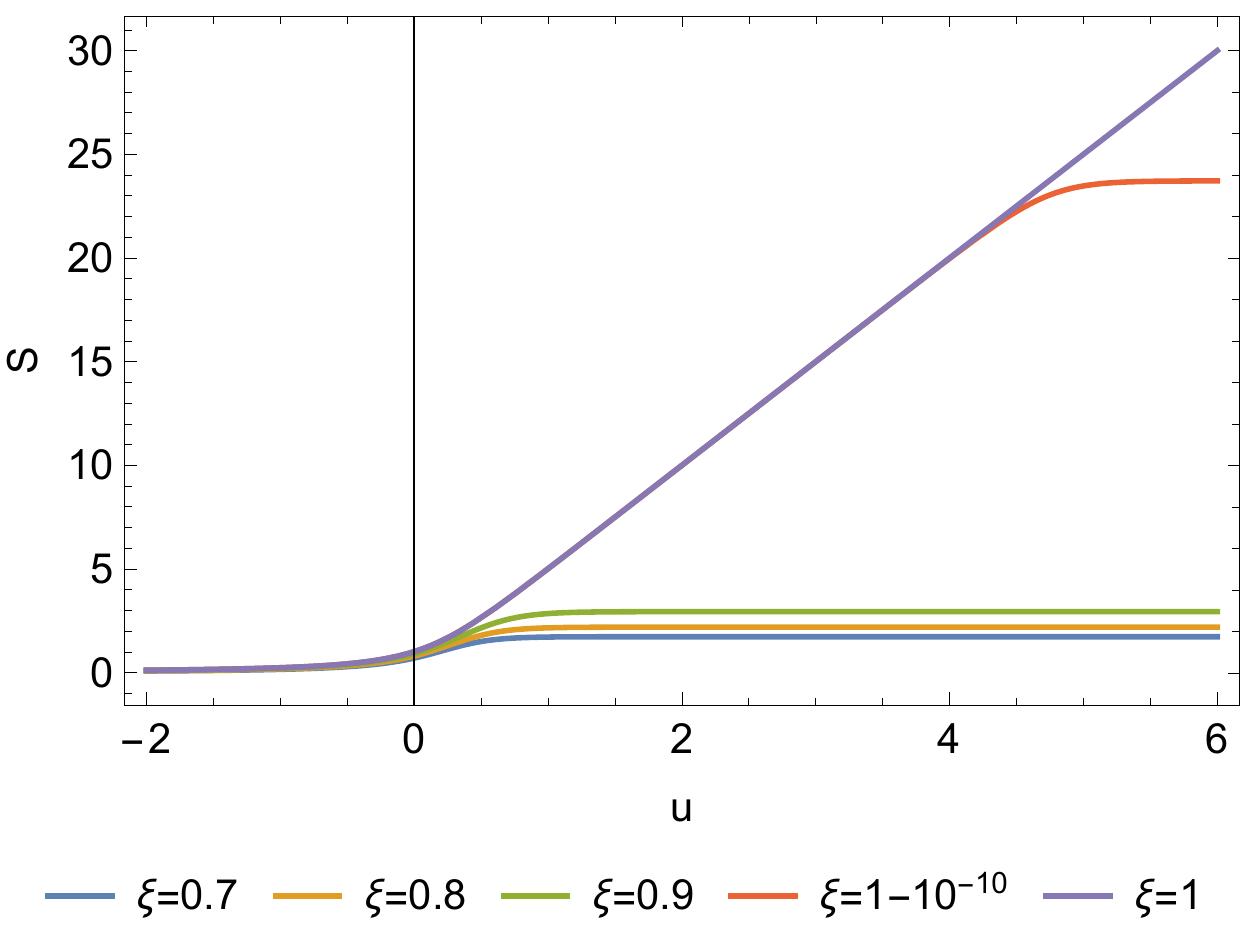}
    \caption{Plot of von Neumann entanglement entropy, $S(u) = -(1/12) \log p'(u)$, for the class of mirror trajectories Eq.~(\ref{eq:traj}) with final drifting speeds $\xi$. Notice the asymptotic divergent entropy for the horizon case $\xi =1$, which is in qualitative agreement with results of Figure \ref{fig: ccrvd}, suggesting non-unitary evolution -information loss in the horizon system is responsible for the divergence in concurrence.}
    \label{fig: entropy}
\end{figure}

 Alternatively, we could consider   the von Neumann entanglement entropy, $S(u)$, which in the present case is given by $S(u) = -(1/12) \log p'(u)$ \cite{Bianchi:2014qua}. Roughly speaking, this measures the amount of entanglement between the two spacetime regions lying respectively before and after the null-line $u$. A plot of $S(u)$ for the different trajectories is given in Fig. \ref{fig: entropy} (see also \cite{Good:2016atu}). For the $\xi=1$ trajectory, there is information loss that can be characterized by a divergent entanglement. A diverging entanglement entropy has also been observed in null-shell collapse to a black hole in \cite{Bianchi:2014bma}. There, the linear increase was interpreted as being due to a constant rate of entanglement entropy production by the black hole. As mentioned in the introduction, the UDW model serves as an operational way of measuring the amount of entanglement present in the vacuum. It is thus satisfying to see an agreement in the qualitative behaviour between the concurrence of the detectors and the entanglement entropy between different spacetime regions.  This correspondence at least suggests that the ever-increasing concurrence that appears for the horizon mirror can be intuitively thought of as a direct result of loss of information in the system. Of course, further case studies are necessary to see if this agreement is a mere coincidence.

\subsection{Negative stress energy}\label{nef}
The trajectories with $\xi<1$ of Eq.~\eqref{eq:traj} have no horizons. These are particularly interesting because they give rise to negative stress-energy in certain spacetime regions. In generic mirror spacetimes, the stress-energy tensor is given by the Schwarzian derivative
\begin{eqnarray}
\label{energyflux}
   F \equiv T_{uu}& = &-\frac{1}{24\pi}\{p(u),u\} \\
    & \equiv & -\frac{1}{24\pi}\left[\frac{p'''(u)}{p'(u)}-\frac{3}{2}\left(\frac{p''(u)}{p'(u)}\right)^2\right]\,.
\end{eqnarray}
A sum rule in proper time demonstrates that information preservation implies negative energy flux emission.  The radiation emitted is calculated by the above Davies-Fulling-Unruh formula where the reflecting boundary trajectory has  rapidity $w(t) = \tanh^{-1} \dot{x}_m(t)$ or $w(u) = \frac{1}{2}\log p'(u) = \log \tau_m'(u)$ \cite{Good:2017ddq}.  Expressing the motion in terms of $\tau_m$, the proper time of the mirror, the radiation flux $F$, is simply
\be 12\pi F(\tau_m)= -w''(\tau_m)e^{2w(\tau_m)},\ee
demonstrating that jerking toward an observer at $\mathscr{I}^+$, with $+w''(\tau_m)$, yields negative energy flux.  Integrating gives  
\be 12\pi\int^\infty_{-\infty} d\tau_m\,e^{-2w(\tau_m)} F(\tau_m)=-\left.w'\right|^{+\infty}_{-\infty}.\ee
Since our mirror Eq.~\eqref{eq:traj} always moves slower than the speed of light (causality), even asymptotically, then $w$ becomes constant for $\tau_m \rightarrow \pm \infty$, and we obtain a sum rule,
\be \int^\infty_{-\infty} d\tau_m\,e^{-2w(\tau_m)} F(\tau_m) = 0.\label{sum}\ee
On the general principle of a universal asymptotic speed limit that remains time-like (as $\tau_m\rightarrow \infty$, then $w\neq \infty$), asymptotic horizonless mirrors will therefore radiate a negative energy flux.  Through the information-dynamics relationship \cite{Good:2016atu,Chen:2017lum,Good:2020nmz,Bianchi:2014qua} $w = -6S$, the time-like restriction corresponds to a pure state, i.e.  the entanglement entropy never diverges and  unitary evolution implies negative energy flux.  This has a correspondence with  a black hole system in terms of a transient increase in black hole mass during evaporation, insofar as a spherically symmetric collapsing matter distribution can be described by a two-dimensional massless conformal field theory, neglecting backscattering, via the s-wave sector of the Hawking radiation carrying the bulk of the radiated energy (see e.g. \cite{purity}).

\begin{figure}
    \centering
    \includegraphics[scale=0.6]{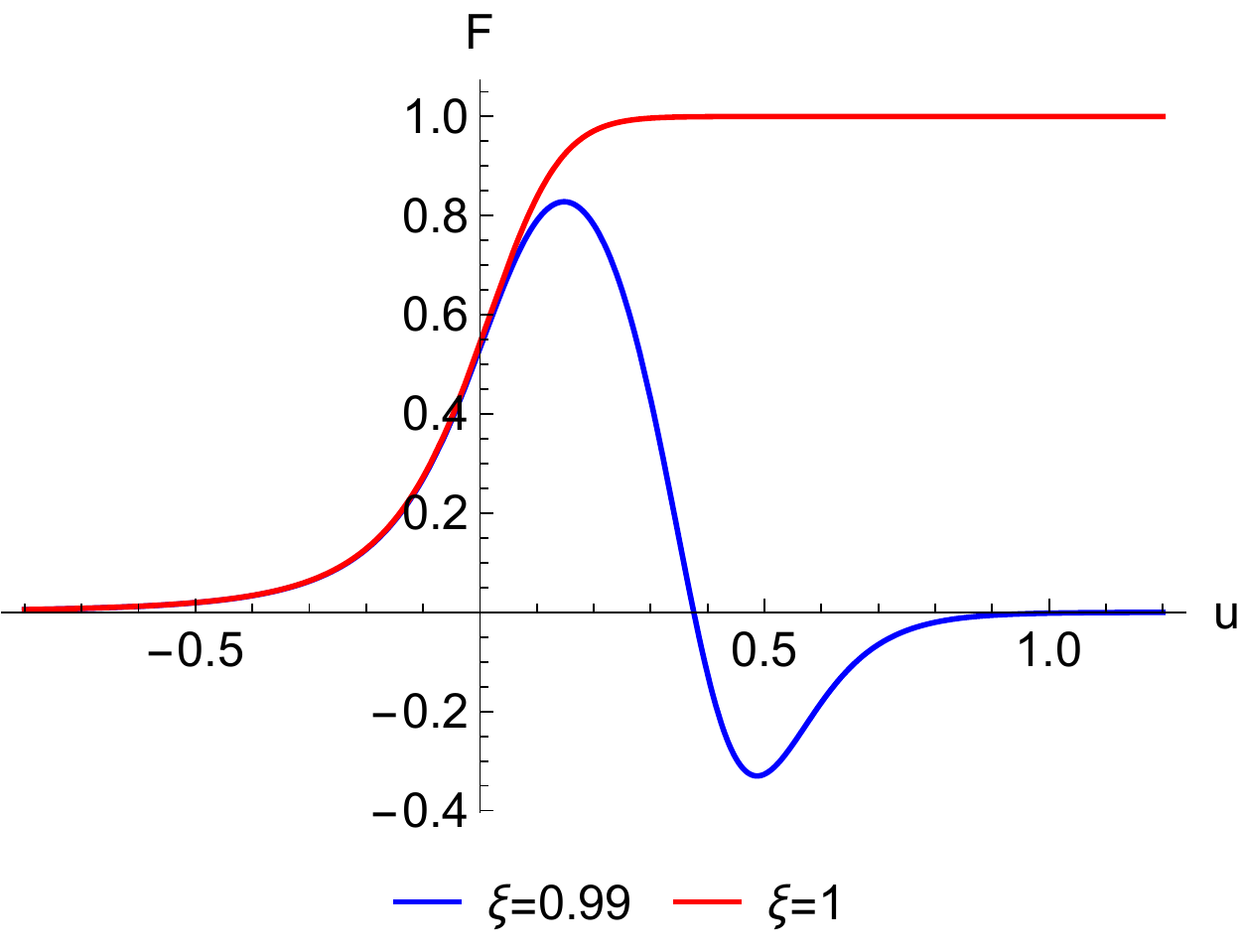}
    \caption{Plot of energy flux, Eq.~(\ref{energyflux}), $F$, against $u$, for the trajectory Eq.~\eqref{eq:traj} (blue) with $\kappa=\sqrt{48\pi}$ and $\xi=0.99$. The scale of the system, $\kappa$, is chosen so that thermal emission is at $F=1$ and a comparison with positive energy flux of the BHC mirror, Eq.~(\ref{bhc}) (red), is illustrated.  Negative energy flux is radiated by the asymptotically inertial motion.}
    \label{fig: stress}
\end{figure}

A plot of flux versus delayed time $u$, of the current trajectory is shown in Fig. \ref{fig: stress} for $\kappa=\sqrt{48\pi},\,\xi=0.99$, which explicitly demonstrates this negative energy flux.

\begin{figure}
    \centering
    \includegraphics[scale=0.45]{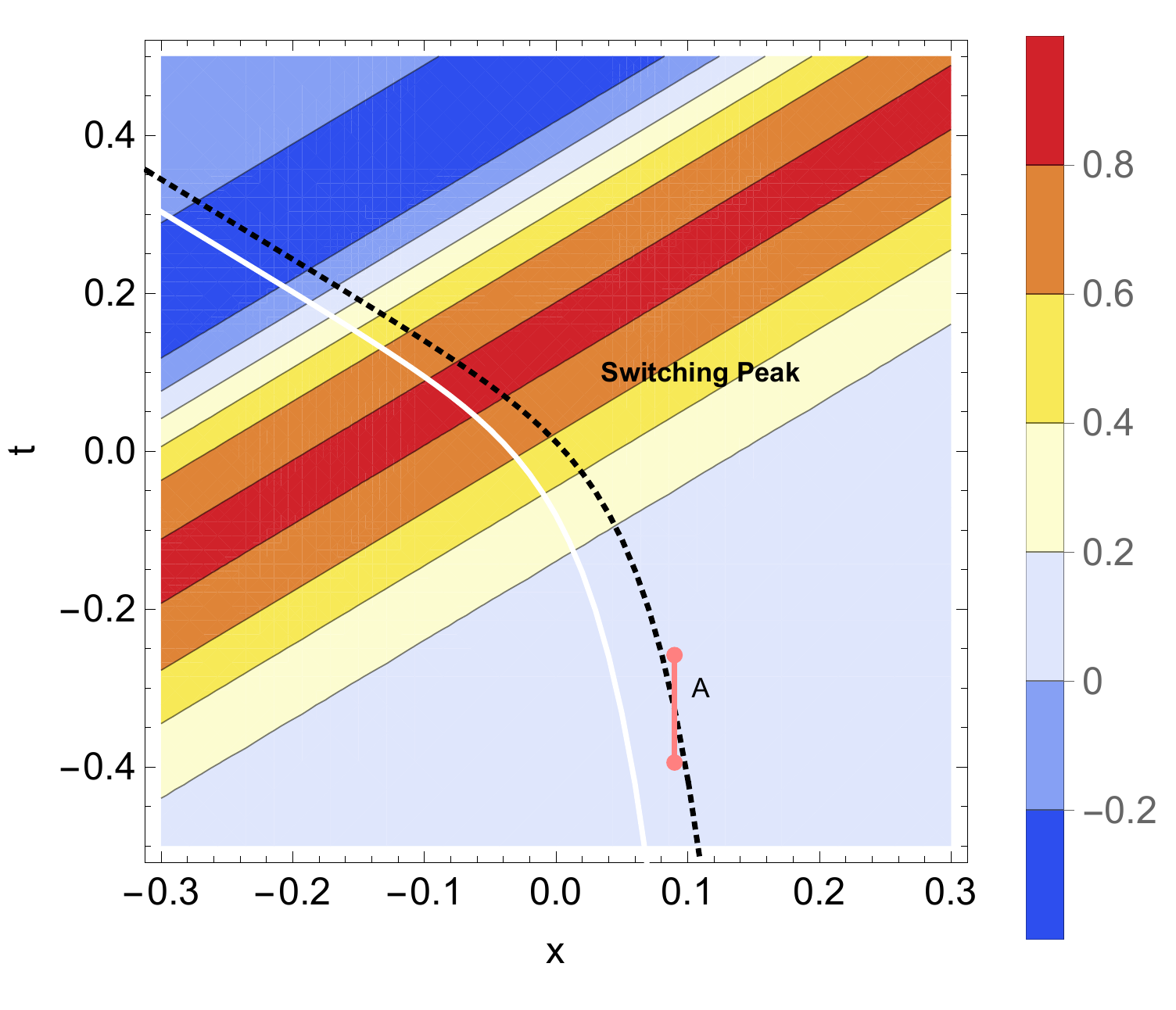}
    \includegraphics[scale=0.5]{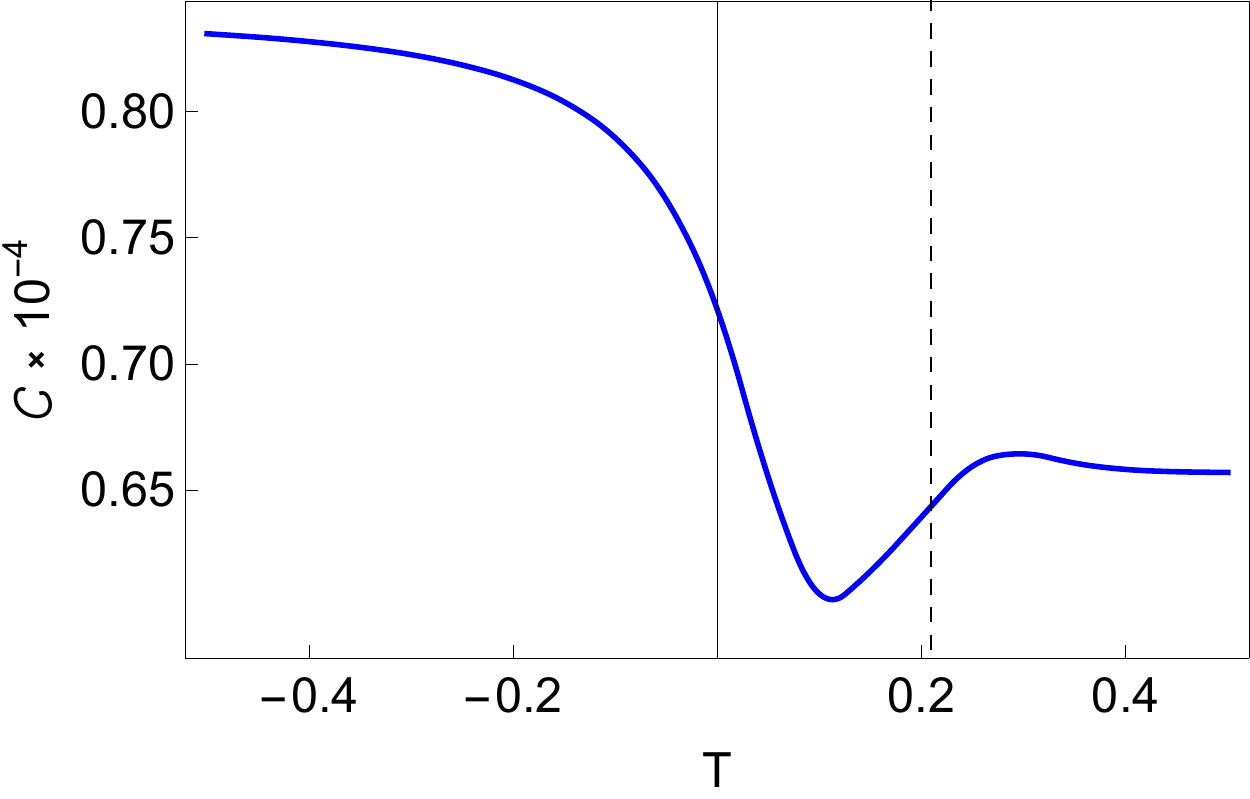}
    \caption{Insensitivity of concurrence to energy density sign change. In the mirror spacetime, the energy density $T_{tt} = T_{uu}$ depends only on $u$ and is constant on these null lines. This is clearly reflected on the top plot, whose colour density indicates the local energy density in the $t/x$ plane. For the $\xi=0.99$ mirror (white trajectory) used here, the energy density changes from positive to negative at some intermediate $u$. To investigate the impact of the change in sign on entanglement harvesting, we consider a series of static detector pairs, each switching on at different $u$ values. The red bar on the top figure shows an example of the spacetime support of the switching function of detector $A$. The detector settings used here are $\Delta \tau = \pi/\eta = 0.136, \Delta x = 0.05, \Omega = 50, \kappa = \sqrt{48\pi}$, and $d_A = 0.04$ (black dashed). With this $d_A$ setting, detector $A$ passes from positive to negative energy density at $T \approx 0.2095$. 
    The concurrence $\mathcal{C}$ against $T$ plot is shown on the bottom. We see that the concurrence varies smoothly across $T \approx 0.2095$, indicating that the concurrence is not directly affected by a change in the sign of energy density.}
    \label{fig: cvs}
\end{figure}

\begin{figure}
    \centering
    \includegraphics[scale=0.39]{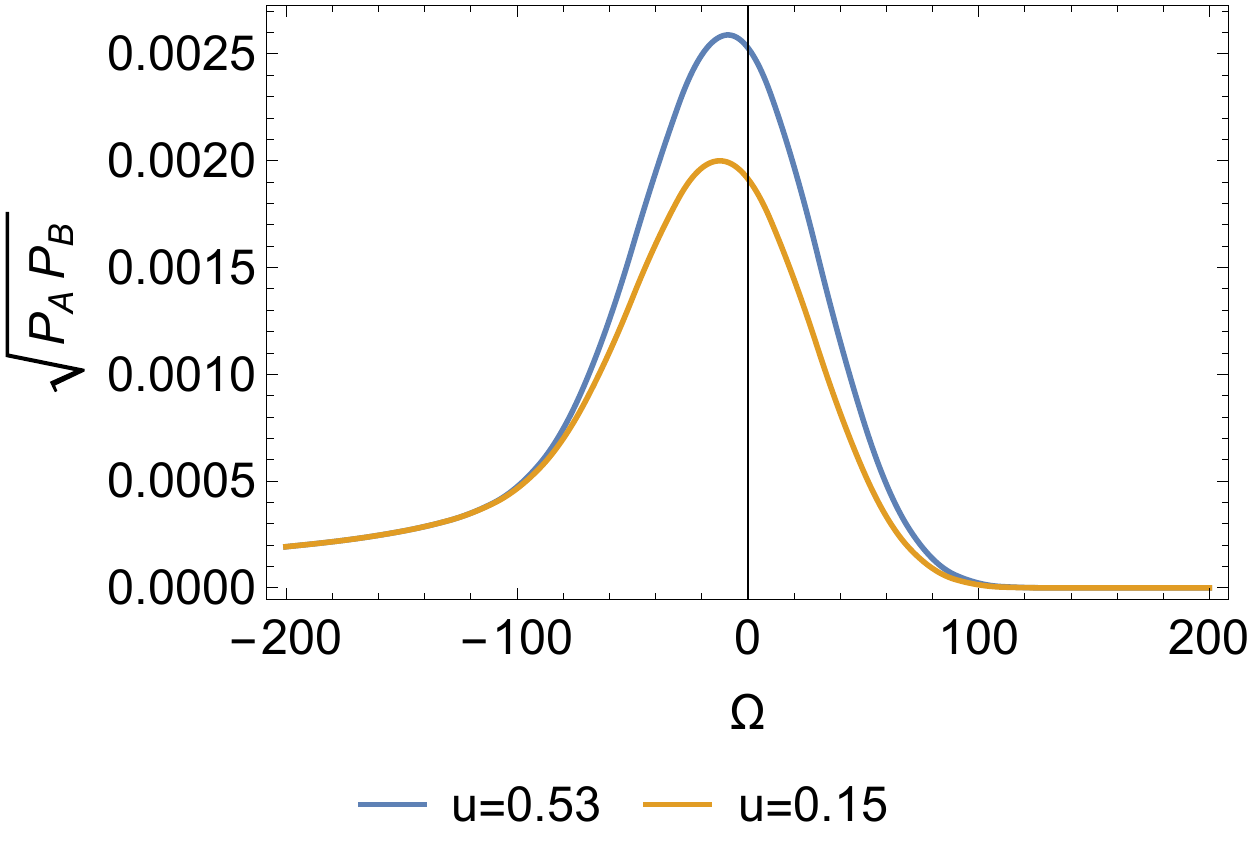}
    \includegraphics[scale=0.39]{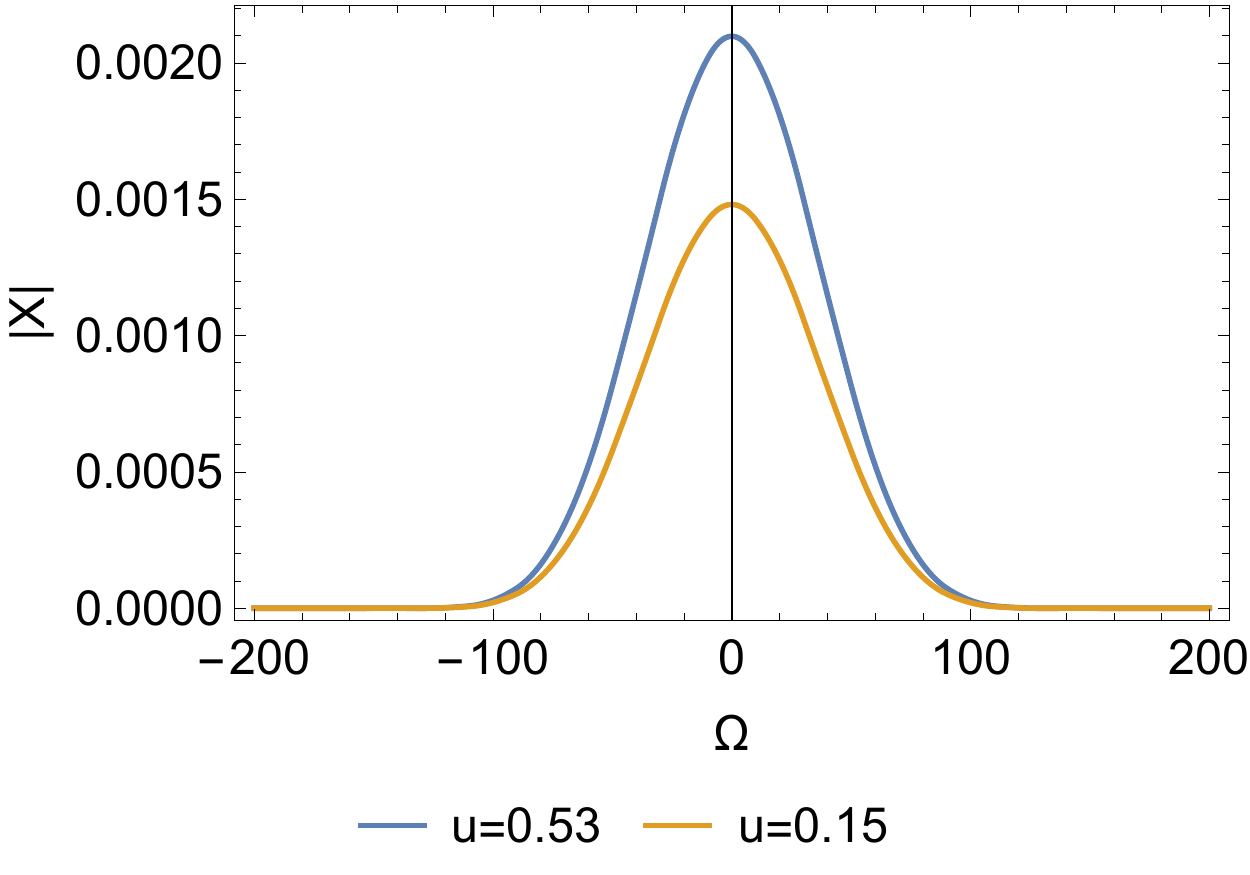}
    \includegraphics[scale=0.37]{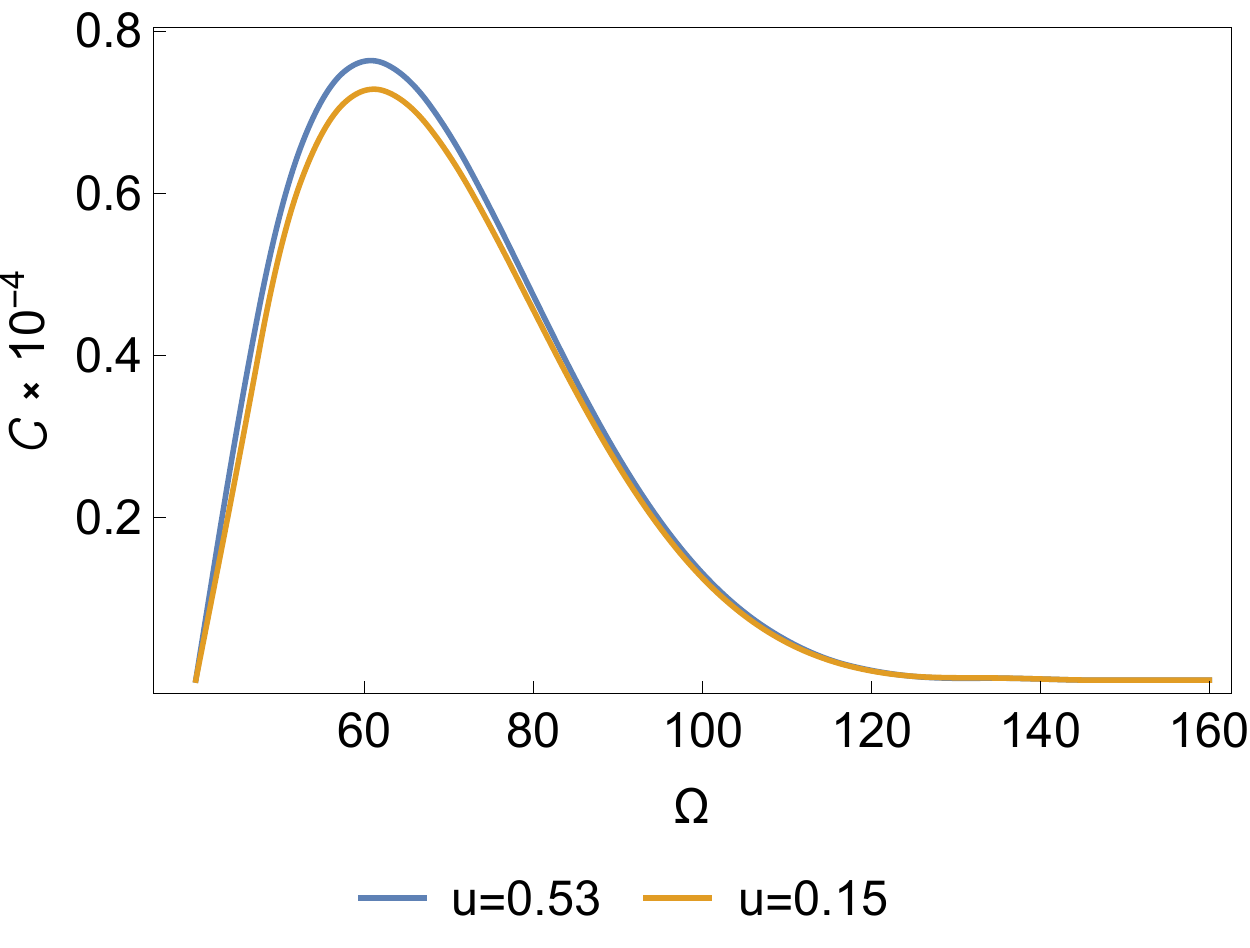}
    \includegraphics[scale=0.39]{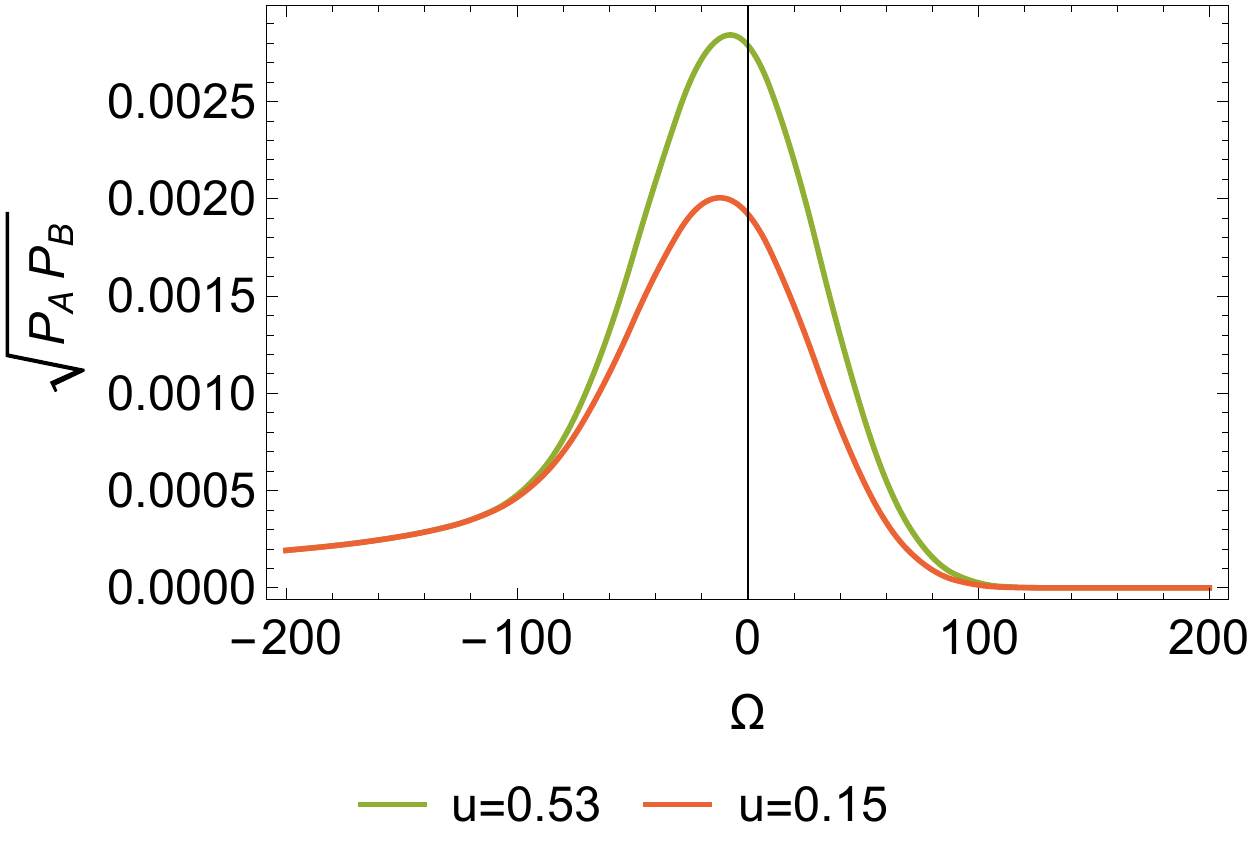}
    \includegraphics[scale=0.39]{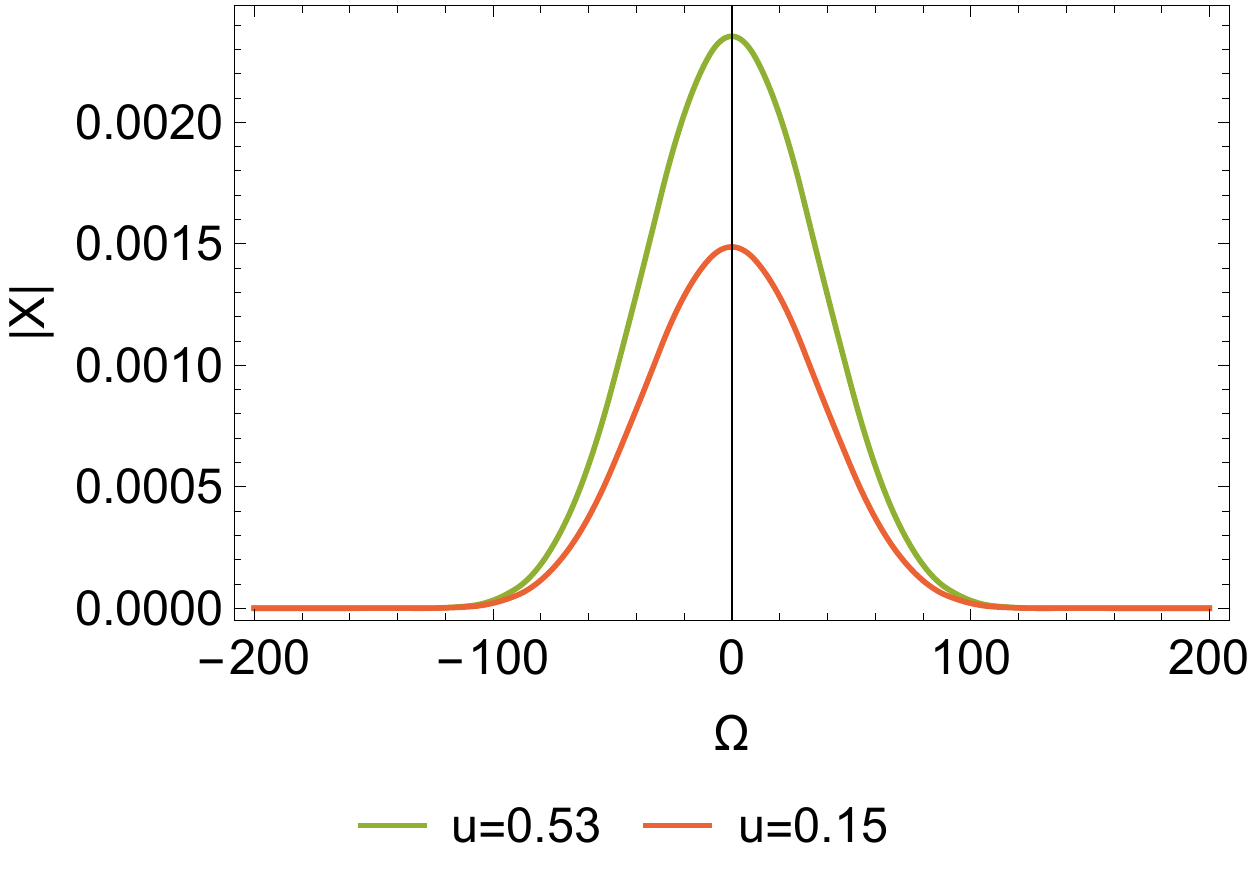}
    \includegraphics[scale=0.37]{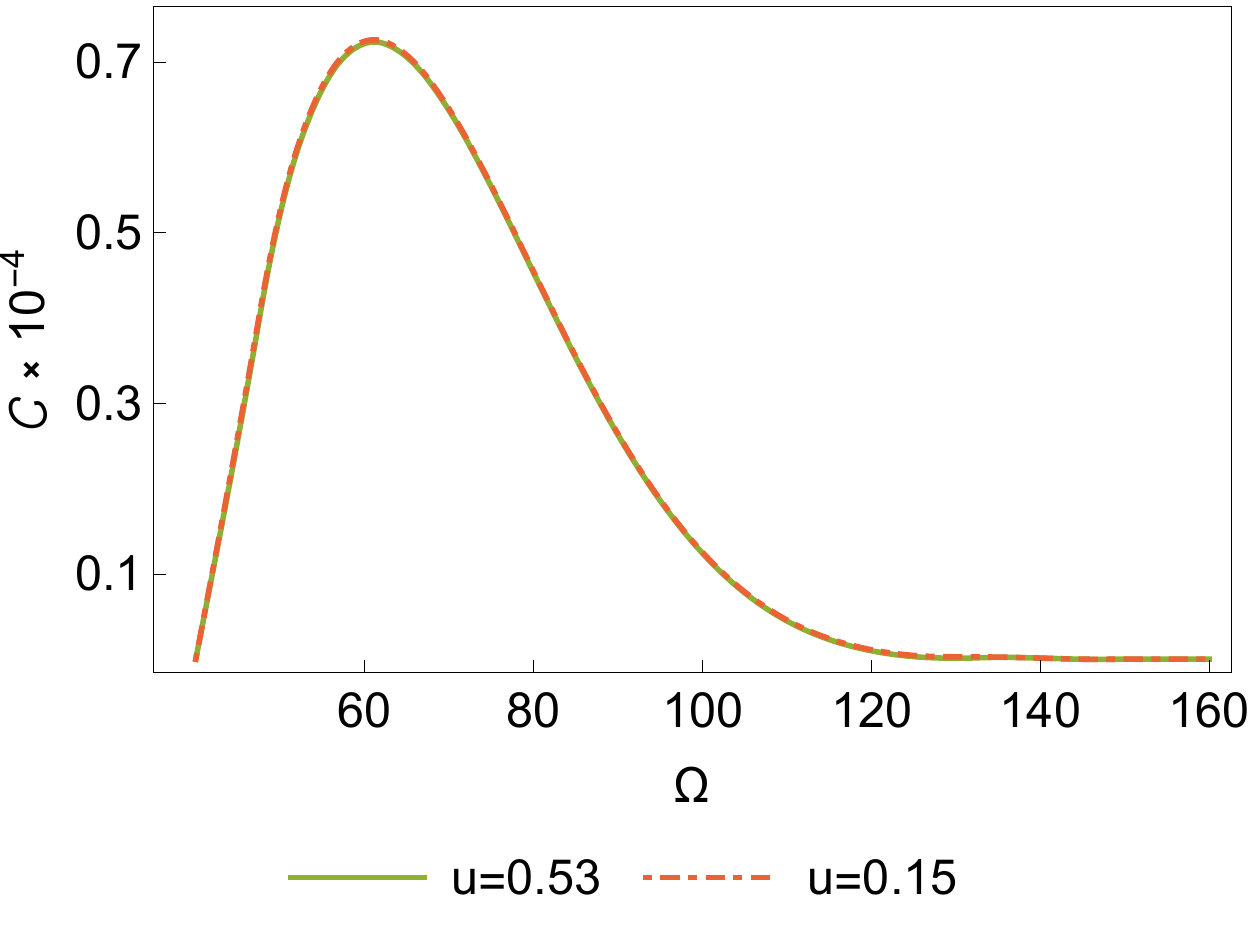}
    \caption{Insensitivity of detector excitation spectrum to the sign of the  energy density. While the local energy density for $\xi=0.99$ mirror can be negative, that of the $\xi=1$ mirror is always positive (see Fig. \ref{fig: stress}). It is known that the transition rate of detectors exhibit a thermal spectrum at the temperature corresponding to that of the surrounding thermal radiation in the case of black hole radiation and Unruh effect. Though we are unable to directly compute the spectrum of the detectors due to the numeric nature of this work, we can nonetheless plot both $P_j$ and $|X|$ against $\Omega$ numerically to see how both depend on the sign of the energy density. The two curves on each of these plots are obtained by respectively placing detector $A$ at $u=0.15,0.53$, which experiences a positive and negative energy density (for the $\xi=0.99$ mirror). The detector settings are $\Delta \tau = 0.136, \Delta x = 0.05, \Omega = 50, \kappa = \sqrt{48\pi}, x_A = 0$. The results for $\xi=0.99$ (\textbf{top}) mirror shows that regardless of the sign of the energy density, the noise ($\sqrt{P_A P_B}$), correlation and concurrence plots assume similar shapes. The differences in the heights of the curves are not a result of the difference in sign of the energy density, since the same is also observed for the $\xi=1$ (\textbf{bottom}) mirror (energy density positive everywhere).}
    \label{fig: sign}
\end{figure}

It is known that the experience of a particle detector may not reflect the energy density given by the renormalised stress energy tensor except in special cases such as in black hole radiation and in the Unruh effect \cite{Birrell:1982ix}. In these cases, the response of particle detectors are given by a thermal spectrum. In this section, we investigate whether the detector response is sensitive to the sign of the energy density. The results are shown in Fig. \ref{fig: cvs} and Fig. \ref{fig: sign}.

As shown in Fig. \ref{fig: stress}, the energy density due to a $\xi<1$ mirror becomes negative at some value of $u$. Therefore, we consider a series of static detector pairs with each pair switching on at different $u$ values. This is done in Fig. \ref{fig: cvs}. The peaks of the switching function of the A detectors are indicated by the black dashed lines on the top figure. The contours on this figure scales according to the local energy density. The plot of $\mathcal{C}$ against $T$ is shown on the bottom figure. The energy density experienced by detector A changes sign at around $T=0.2095$, indicated on the figure by a vertical dashed line. We see that the concurrence nonetheless varies smoothly across this line. 

Instead of restricting to one mirror trajectory, we can also compare the $\xi<1$ mirror with the $\xi=1$ mirror. The latter radiates a thermal spectrum at late times, and the energy density remains positive for all times. This comparison is made in Fig. \ref{fig: sign}, where we looked at the $\sqrt{P_A P_B}$ against $\Omega$ and $\mathcal{C}$ against $\Omega$ graphs. Intuitively, we might expect negative energy density to induce more de-excitation of particle detectors, resulting in a shift in the peak of the $\sqrt{P_A P_B}$ against $\Omega$ towards negative $\Omega$ ($P_A$ is then interpreted as de-excitation probability). However, we see that this is not the case from the left figures. In fact, both the blue ($u=0.53$, negative $T_{uu}$) and yellow ($u=0.15$, positive $T_{uu}$) curves peak at around the same $\Omega$ and share the same overall shape despite a difference in magnitude that may be attributed to a difference in $d_A$ and instantaneous mirror velocity. We therefore conclude that the negative local energy densities in $(1+1)$D mirror spacetimes cannot be detected from the entanglement harvested by UDW detectors.

\section{Conclusions}\label{conc}

We have addressed  the question raised in Sec.~\ref{intro} by investigating the effect of horizons on entanglement harvesting.  Our approach was to merge the model of entangled Unruh-Dewitt detectors in spacetime~\cite{Smith2016topology} and properties of an accelerating mirror spacetime with or without a horizon~\cite{Good:2016atu}. We presented the difference between horizonless mirrors and horizon mirrors (black hole collapse mirrors) in two main respects: concurrence of harvested entanglement and sensitivity of the detector to the sign of radiated energy flux.

We find that concurrence can distinguish between the global property of a dynamic spacetime containing a horizon and one without. However, the effect is subtle and harvesting without horizons does not dramatically affect entanglement. The sudden death of entanglement   occurs for both horizon mirrors and horizonless mirrors, as we depict  in $\mathcal{C}-d_A$ plots of Fig. \ref{fig: ccrvd}.   However for horizon mirrors   concurrence  at small $d_A$ ``revives" as time increases.  The most striking difference is illustrated in Fig. \ref{fig: ccrvd}: concurrence for horizonless mirrors asymptotes to finite values at large $T$, but for horizon mirrors  concurrence   evidently has  no asymptote.  

Moreover, we find that local energy flux has no sudden direct consequence on entanglement. As is shown in Fig. \ref{fig: stress}, negative energy is radiated for a sub-light speed trajectory, and we have demonstrated that NEF is present for all asymptotically time-like mirrors via a sum rule in proper time, Eq.~(\ref{sum}).  For asymptotically null trajectories, $w\rightarrow \infty$, there is no negative energy radiated. Nevertheless, concurrence is insensitive to the appearance of negative energy flux, which is illustrated in the results of Fig. \ref{fig: cvs}.

 
It will be interesting for further studies to find out why concurrence in the presence of horizons exhibits a death zone that can ``revive" at large $T$, which may depend on properties of the horizon. Likewise, it will be even more interesting to see what features of this study are preserved in actual gravitational collapse.

\appendix

\section{Constant Light Speed Boundary}\label{light}

The light speed mirror produces no particles.  This is most easily seen by computing the beta coefficient \cite{Good:2016atu} in the null-coordinate $u$:
\begin{equation} \beta_{\omega\omega'} = \frac{1}{4\pi\sqrt{\omega\omega'}}\int_{-\infty}^{\infty}du\; e^{-i \omega u-i\omega' p}(\omega'p' - \omega)
\end{equation}
with $p(u) = \frac{1-v_0}{1+v_0}u$ \cite{Good:2013lca}, where $v_0$ is the drift speed of any constant velocity mirror, in our light speed case, $v_0 = 1$, so that $p(u) = 0$. Therefore, our expression for beta becomes
\begin{equation} 
\beta_{\omega\omega'} \sim \int_{-\infty}^{\infty}du\; e^{-i \omega u} =  2\pi \delta(\omega) = 0,
\end{equation}
where in the last step, positive frequency $\omega> 0$ takes care of the Dirac delta.  

\section{Asymptotic concurrence values for $\xi<1$ mirrors}
\label{asymptotic}
Recall that in order to compute the concurrence, we need to compute the probability $P$ as well as a correlation $X$. To explain the asymptotic behaviour, we only need to look at $P$ as the analysis for $X$ will be similar. We have two ways to measure the probability ``at time $t=T$'': we can either directly impose $\chi(t)=\cos^4(\eta(t-T))$, or shift the trajectory down by $T$ units by setting $v_H=-T$. These two methods are physically equivalent and yield the same results, but we will use the second one to explain the asymptotic behaviour in Fig. \ref{fig: ccrvd}.

Using the second method, the dependence on $T$ will appear in the Wightman function, which is comprised of four pieces of logarithmic functions \cite{Birrell:1982ix}:
\begin{equation}
    \label{eq:wightman}
        W(t,x_d;t',x_d) = -\frac{1}{4\pi}\log\Bigg[\frac{(p(u)-p(u')-i\epsilon) (v-v'-i\epsilon)}{(p(u)-v'-i\epsilon)(v-p(u')-i\epsilon)}\Bigg]\,,
    \end{equation}
where $v=p(u)$ parametrises the trajectory of the mirror. Let us look at the first piece given by $\log\big[p(t-x_d)-p(t'-x_d)-i\epsilon\big]$ (recall that the Wightman function appearing in Eq.~(\ref{eq: p}) is evaluated along the detector trajectory). Since we are placing the detector at some fixed $d_A$ away from the mirror, where the distance is measured at time $t=T$, we set $x_d=x_m(0,T)+d_A$. Using the trajectory Eq.~\eqref{eq:traj} with $v_H=-T$ we have

\begin{align}
&p(t-x_d)-p(t'-x_d)\nonumber \\
     =&u+\frac{\xi}{\kappa}\log\bigg[\frac{1+\xi}{2}W\big(\frac{2}{1+\xi}e^{\frac{2\kappa(-T-u)}{1+\xi}}\big)\bigg]-u'-\frac{\xi}{\kappa}\log\bigg[\frac{1+\xi}{2}W\big(\frac{2}{1+\xi}e^{\frac{2\kappa(-T-u')}{1+\xi}}\big)\bigg]\nonumber\\
     =&t-t'+\frac{\xi}{\kappa}\log\bigg[W\big(\frac{2}{1+\xi}e^{\frac{2\kappa(-T-u)}{1+\xi}}\big)\bigg]-\frac{\xi}{\kappa}\log\bigg[W\big(\frac{2}{1+\xi}e^{\frac{2\kappa(-T-u')}{1+\xi}}\big)\bigg]\nonumber\\
     =&t-t'+\frac{\xi}{\kappa}\log\bigg[\frac{2}{1+\xi}e^{\frac{2\kappa(-T-u)}{1+\xi}}\bigg]-\frac{\xi}{\kappa}W\big(\frac{2}{1+\xi}e^{\frac{2\kappa(-T-u)}{1+\xi}}\big)\nonumber\\
     &\quad-\frac{\xi}{\kappa}\log\bigg[\frac{2}{1+\xi}e^{\frac{2\kappa(-T-u')}{1+\xi}}\bigg]+\frac{\xi}{\kappa}W\big(\frac{2}{1+\xi}e^{\frac{2\kappa(-T-u')}{1+\xi}}\big)\nonumber\\
     =&t-t'-\frac{2\xi(T+u)}{1+\xi}-\frac{\xi}{\kappa}W\big(\frac{2}{1+\xi}e^{\frac{2\kappa(-T-u)}{1+\xi}}\big)+\frac{2\xi(T+u')}{1+\xi}+\frac{\xi}{\kappa}W\big(\frac{2}{1+\xi}e^{\frac{2\kappa(-T-u')}{1+\xi}}\big)\,, \label{eq:5}
\end{align}

where in going from the second to the third equality we used the identity $\log [W(x)]=\log[x]-W(x)$. Next, we would like to take the $T\rightarrow \infty$ limit. First, note that $u=t-x_d=t-x_m(T)+d_A$. Meanwhile, 
\begin{equation}
     x_m(T) = -\xi T-\frac{\xi}{2\kappa}W\big(2e^{-2\kappa T}\big)\rightarrow -\xi T
\end{equation}
since $W(0)=0$. Making use of this fact again for the terms involving $W(\cdot)$ in the previous equation, we have
\begin{align}
    p(t-x_d)-p(t'-x_d) &\rightarrow t-t'-\frac{2\xi(T+u)}{1+\xi}+\frac{2\xi(T+u')}{1+\xi}\nonumber\\
    &=t-t'-\frac{2\xi(u-u')}{1+\xi}\nonumber\\
    &=\bigg(1-\frac{2\xi}{1+\xi}\bigg)(t-t')\,.\label{eq:asym}
\end{align}

Hence, $\log\big[p(t-x_d)-p(t'-x_d)-i\epsilon\big]\rightarrow \log[\big(1-\frac{2\xi}{1+\xi}\big)(t-t')-i\epsilon]$. This asymptotic form coincides with the piece contained in the Wightman function of a mirror moving to the left at constant time-like speed $\xi<1$ passing through the origin. The trajectory of such a mirror is $t=-1/\xi x$, or equivalently, \begin{equation}
    v=(1-\frac{2\xi}{1+\xi})u\equiv p_c(u)\,.
\end{equation} 
Hence as promised, the corresponding piece $\log\big[p_c(t-x_d)-p_c(t'-x_d)-i\epsilon\big]=\log[\big(1-\frac{2\xi}{1+\xi}\big)(t-t')-i\epsilon]$ in the Wightman function. Repeating this for each of the three remaining pieces in eq.~\eqref{eq:wightman}, we will find that \textit{at large $T$, the Wightman function approaches that of the constant speed mirror}. 

We have thus almost successfully explained the asymptotic value of concurrence of the time-like mirrors in Fig. \ref{fig: ccrvd}. What is left is to show that the same happens for the correlation $X$ term, but it is almost completely the same so we will omit the calcuation.

Finally, we attempt to investigate whether the concurrence with a light-like mirror asymptote to a finite, non-zero value. For any $\xi<1$, the expression in eq.~\eqref{eq:asym} equates to a finite value whenever $t\neq t'$. This gives a finite $P$ when the UV-regulator is taken to $\epsilon\rightarrow 0$ at the end. However when $\xi=1$, the expression in eq.~\eqref{eq:asym} is identically zero for all $t\,,t'$ values, giving $ \log\big[p(t-x_d)-p(t'-x_d)-i\epsilon\big] \rightarrow \log(-i\epsilon)$ which diverges in the limit $\epsilon\rightarrow 0$. Due to this behavior of the Wightman function, an asymptotic value of $P$ for large $T$ does not exist for $\xi=1$. To investigate how the divergence occurs when $\xi=1$ at large $T$, we expand the terms involving the $W(\cdot)$ functions in eq.~\eqref{eq:5} to subleading order in $e^{-\kappa T}$:

\begin{align}
   &\quad-\frac{1}{\kappa}W\big(e^{\kappa(-T-u)}\big)+\frac{1}{\kappa}W\big(e^{\kappa(-T-u')}\big)\nonumber\\ &= -\frac{1}{\kappa} W\big(e^{-\kappa T-\kappa(t+d_A+T+\frac{1}{2\kappa}W\big(2e^{-2\kappa T}\big)}\big)+\frac{1}{\kappa}W\big(e^{-\kappa T-\kappa(t'+d_A+T+\frac{1}{2\kappa}W\big(2e^{-2\kappa T}\big))}\big)\nonumber
   \\&\rightarrow \frac{e^{-2 T\kappa}}{\kappa}(e^{\kappa(d_A-t')}-e^{\kappa(d_A-t)})+O(e^{-4T\kappa})\,.
  \end{align}

\begin{figure}[t]
    \centering
    \includegraphics[scale=0.5]{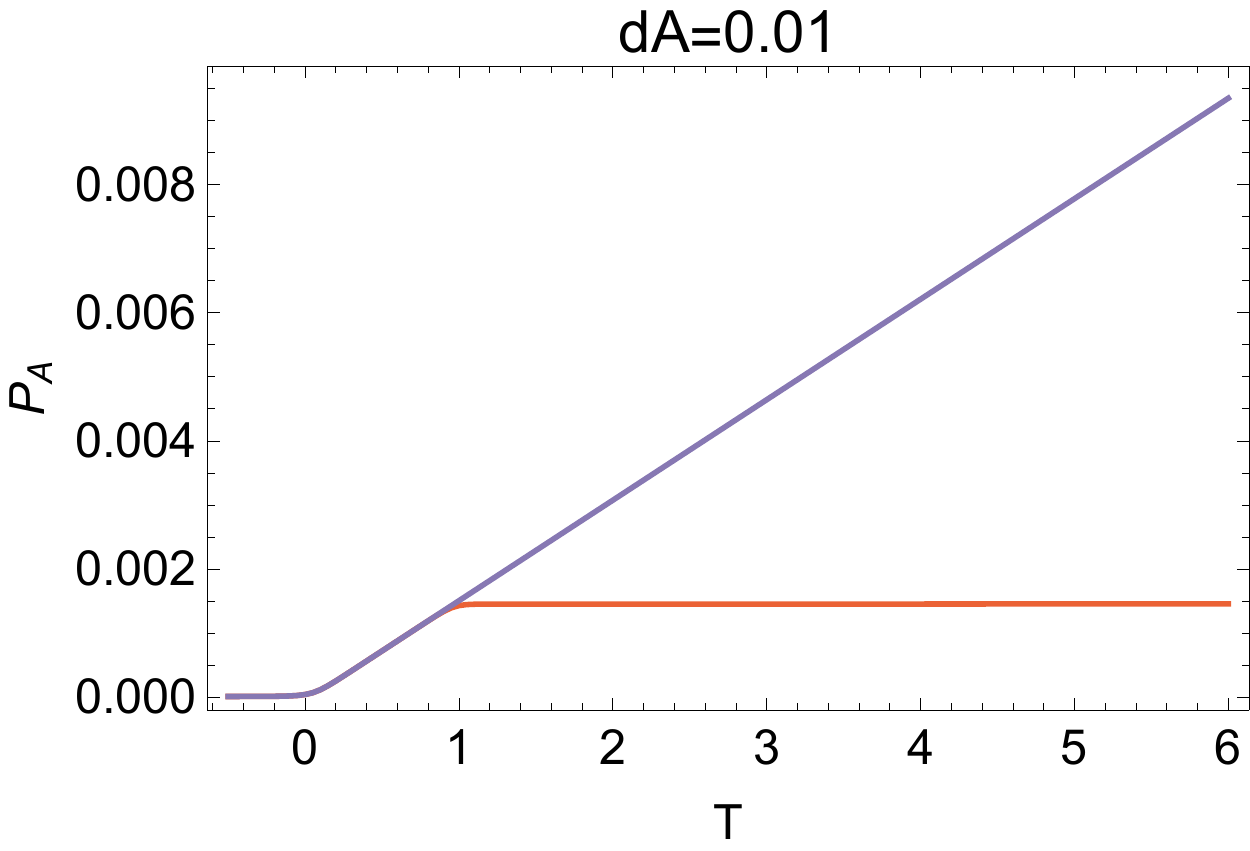} \,
    \includegraphics[scale=0.48]{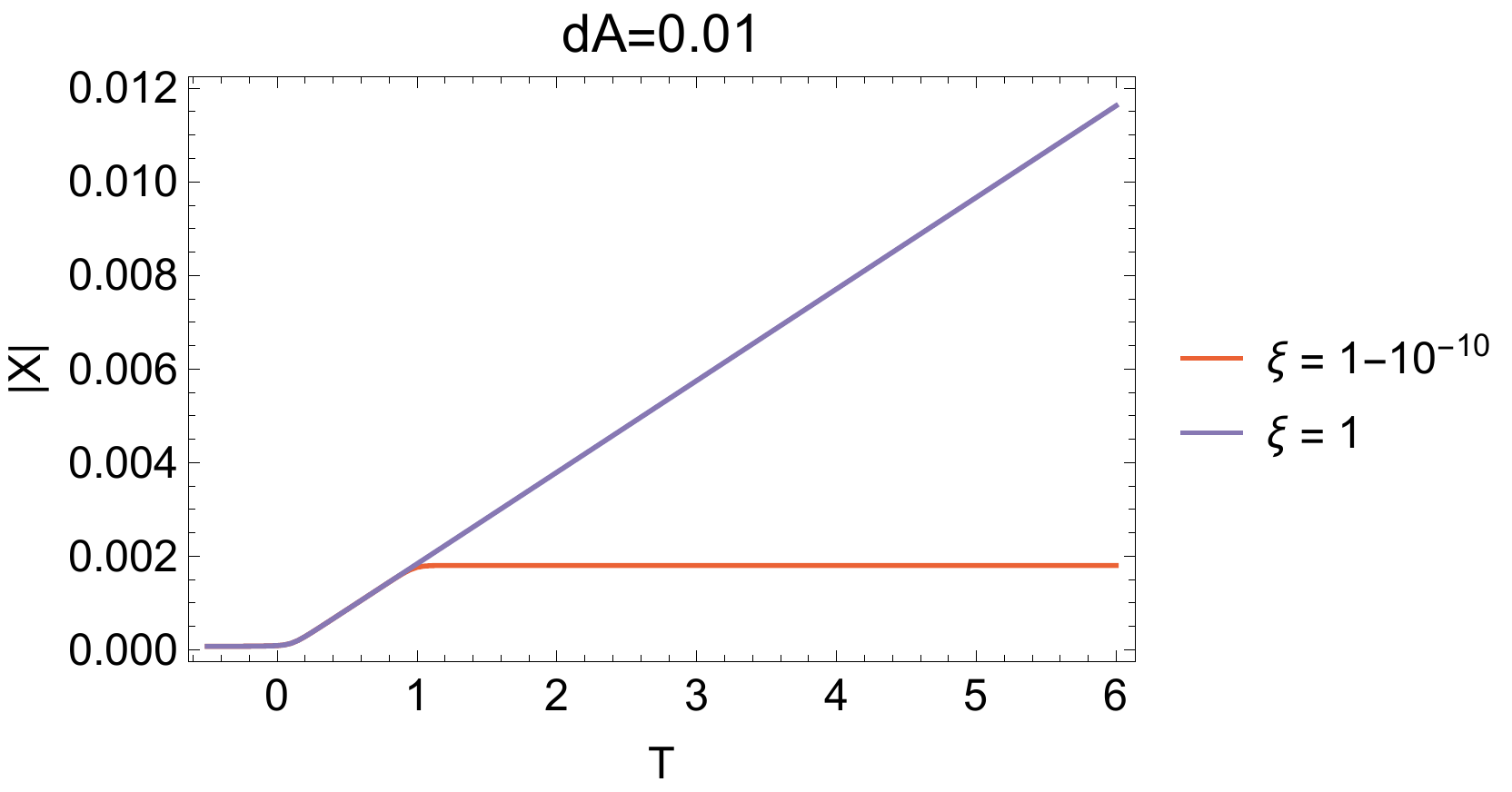}
    \caption{Linear increase in $P$ and $|X|$ for $\xi=1$. The parameters used here are $\kappa = \sqrt{48\pi}$, $\eta = 23$, $\Omega = 50$, $d_A = 0.01$ and $\Delta x = 0.05$.}
    \label{fig: px}
\end{figure}

  Hence for the asymptotically light-like mirror, we have $ \log\big[p(t-x_d)-p(t'-x_d)-i\epsilon\big] \rightarrow \log(\frac{e^{-2 T\kappa}}{\kappa}(e^{\kappa(d_A-t')}-e^{\kappa(d_A-t)})-i\epsilon)=-2\kappa T+\log(\frac{1}{\kappa}(e^{\kappa(d_A-t')}-e^{\kappa(d_A-t)})-i\epsilon)$ after a rescaling of the small parameter $\epsilon$, which blows up at large $T$. An example of the plot of $P_A$ and $|X|$ against $T$ is shown in Fig. \ref{fig: px}.

\section{Numerical convergence}
\label{numerical convergence}

The probability $P$ and correlation $X$ terms as given in Eqs.~\eqref{eq: p} and ~\eqref{eq: x} can be computed numerically. In this section, we briefly comment on the numerical methods adopted in the main text by looking at a couple of illustrative examples.

\begin{figure}
    \centering
    \includegraphics[scale=0.5]{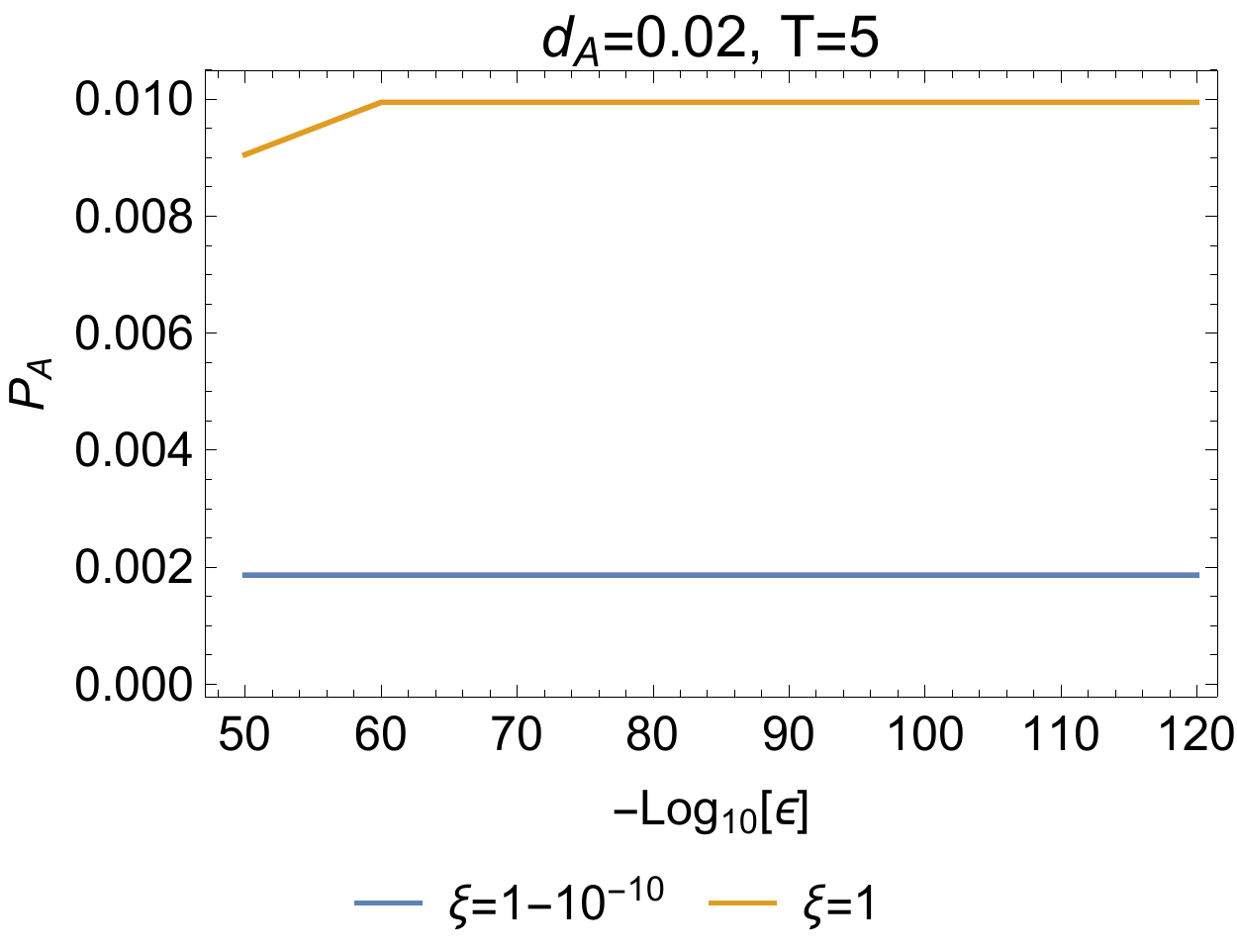}
    \includegraphics[scale=0.5]{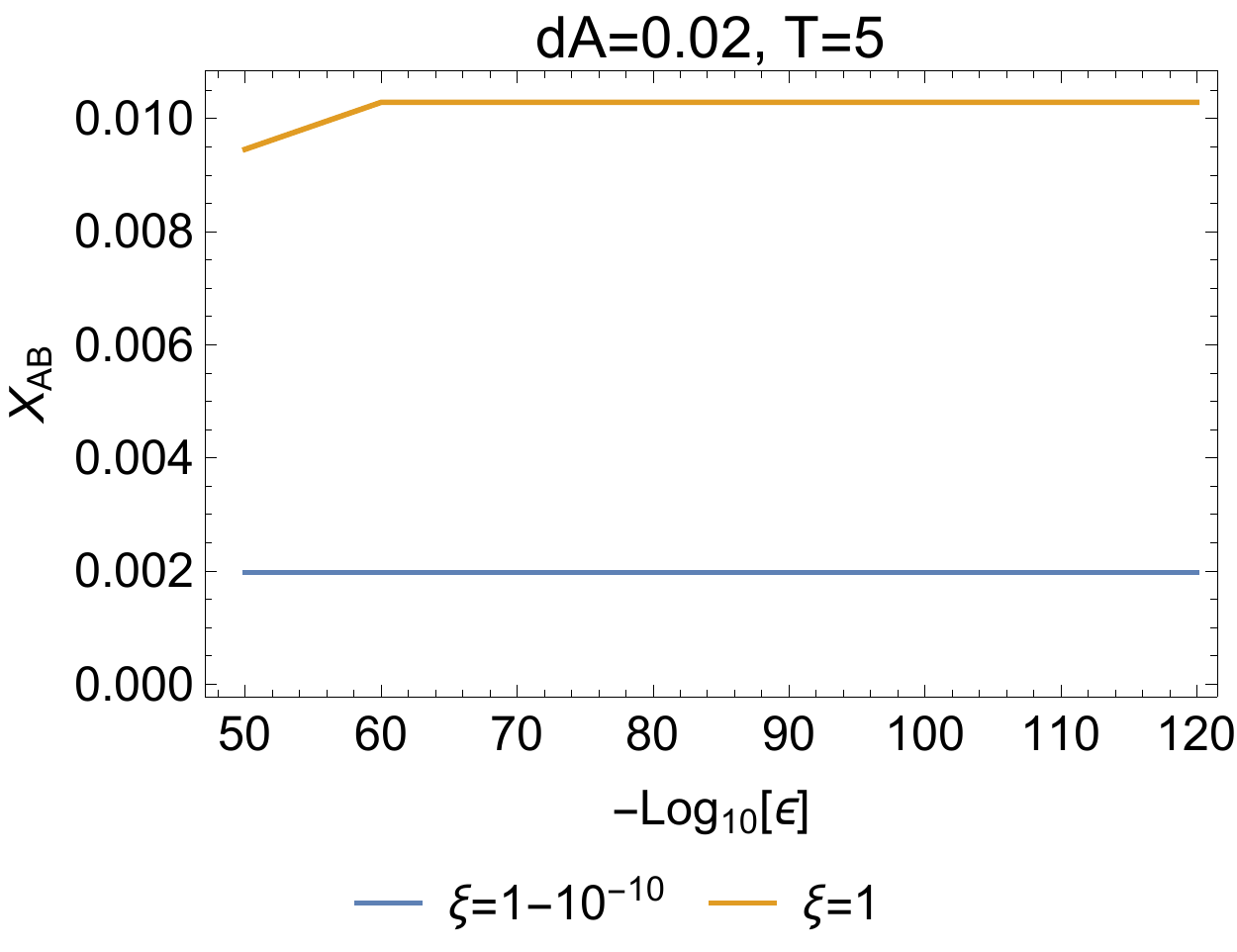}
    \caption{Convergence of $P$ and $|X|$ in $\epsilon$. Shown here are the plots of $P$ and $|X|$ of Eqs.~\eqref{eq: p} and ~\eqref{eq: x} against $\log_{10}\epsilon$, where $\epsilon$ is the uv-regulator appearing in the Wightman function (see Eq.~\eqref{eq:wightman}). The parameters settings are $T=5$ and $d_A = 0.02$, corresponding to an intermediate point in the center figure of Fig. \ref{fig: ccrvd}. Both the $P$ and $|X|$ terms clearly converge for small $\epsilon$.}
    \label{fig:conv}
\end{figure}

First, we note that a \textit{manual} checking of the convergence in the UV-regulator $\epsilon$ (in Eq.~\eqref{eq:wightman}) is necessary. The expressions for $P$ and $X$ are expected to converge for a sufficiently small positive $\epsilon$. Fig. \ref{fig:conv} shows some examples of plots of $P_A$ and $|X|$ against $-\log_{10}\epsilon$, demonstrating the convergence in small $\epsilon$. In the corresponding Fig. \ref{fig: ccrvd} of the main text, the $\epsilon$ value chosen was $\epsilon = 10^{-120}$. From Fig. \ref{fig:conv}, it is clear that both $P$ and $|X|$ have converged to sufficient precision at this $\epsilon$ value. Indeed, the errors due to the UV-regulator can be estimated as $|P_A(\epsilon = 10^{-119})-P_A(\epsilon = 10^{-120})|\approx 10^{-6}$ and $||X(\epsilon = 10^{-119})|-|X(\epsilon = 10^{-120})|| \approx 10^{-12}$ for $\xi = 1-10^{-10}$ and $|P_A(\epsilon = 10^{-119})-P_A(\epsilon = 10^{-120})|\approx 10^{-9}$ and $||X(\epsilon = 10^{-119})|-|X(\epsilon = 10^{-120})|| \approx 10^{-10}$ for $\xi = 1$. These thus give negligible contributions to the concurrence which is $\sim 10^{-3}$. Convergence in $\epsilon$ has been checked for all figures in the main text.

Our numerical integration was implemented using the ``Nintegrate'' function of Mathematica. All settings are set as default, i.e., the ``method'' is ``GlobalAdaptive'' and ``WorkingPrecision'' (without specifying explicitly, by default) is ``MachinePrecision''. However to speed up the computation, we have switched off the ``OscillatorySelection'' option. As a check of the reliability of the numerical results, one can specify a ``WorkingPrecision'' value. For example, setting instead ``WorkingPrecision $\rightarrow 180$'' leads to a $O(10^{-6})$ ($O(10^{-9})$) difference in $P_A$ and $O(10^{-12})$ ($O(10^{-10})$) difference in $|X|$ for $\xi = 1-10^{-10}$ ($\xi = 1$). Once again, this is much smaller than the concurrence, thus demonstrating the numerical reliability of the results.

\acknowledgments
We thank Yen Chin Ong for advice and discussions.
C.Q. is supported in part by the National Natural Science Foundation of China (NSFC) with grant No. 11675164, 11535012 and 11890713.
M.G. is funded by the FY2018-SGP-1-STMM Faculty Development Competitive Research Grant No. 090118FD5350 at Nazarbayev University,  and the state-targeted program ``Center of Excellence for Fundamental and Applied Physics" (BR05236454) by the Ministry of Education and Science of the Republic of Kazakhstan. This work was supported in part by the Natural Sciences and Engineering Research Council of Canada.
\bibliographystyle{JHEP}

\bibliography{main}

\providecommand{\href}[2]{#2}\begingroup\raggedright\begin{thebibliography}{10}

\bibitem{PhysRev.110.965}
D.~Finkelstein, {\it Past-future asymmetry of the gravitational field of a
  point particle},  {\em Phys. Rev.} {\bf 110} (May, 1958) 965--967.

\bibitem{Birrell:1982ix}
N.~Birrell and P.~Davies, {\em {Quantum Fields in Curved Space}}.
\newblock Cambridge Monographs on Mathematical Physics. Cambridge Univ. Press,
  Cambridge, UK, 2, 1984.

\bibitem{Parker:2009uva}
L.~E. Parker and D.~Toms, {\em {Quantum Field Theory in Curved Spacetime}:
  {Quantized Field and Gravity}}.
\newblock Cambridge Monographs on Mathematical Physics. Cambridge University
  Press, 8, 2009.

\bibitem{Davies:1974th}
P.~Davies, {\it {Scalar particle production in Schwarzschild and Rindler
  metrics}},  {\em J. Phys. A} {\bf 8} (1975) 609--616.

\bibitem{Unruh:1976db}
W.~G. Unruh, {\it {Notes on black hole evaporation}},  {\em Phys. Rev.} {\bf
  D14} (1976) 870.

\bibitem{Hawking:1974sw}
S.~Hawking, {\it {Particle Creation by Black Holes}},  {\em Commun. Math.
  Phys.} {\bf 43} (1975) 199--220. [Erratum: Commun.Math.Phys. 46, 206 (1976)].

\bibitem{Mann:2015luq}
R.~B. Mann, {\em {Black Holes: Thermodynamics, Information, and Firewalls}}.
\newblock SpringerBriefs in Physics. Springer, 2015.

\bibitem{Salton:2014jaa}
G.~Salton, R.~B. Mann, and N.~C. Menicucci, {\it {Acceleration-assisted
  entanglement harvesting and rangefinding}},  {\em New J. Phys.} {\bf 17}
  (2015), no.~3 035001, [\href{http://arxiv.org/abs/1408.1395}{{\tt
  arXiv:1408.1395}}].

\bibitem{Pozas-Kerstjens:2015gta}
A.~Pozas-Kerstjens and E.~Martin-Martinez, {\it {Harvesting correlations from
  the quantum vacuum}},  {\em Phys. Rev. D} {\bf 92} (2015), no.~6 064042,
  [\href{http://arxiv.org/abs/1506.03081}{{\tt arXiv:1506.03081}}].

\bibitem{VALENTINI1991321}
A.~Valentini, {\it Non-local correlations in quantum electrodynamics},  {\em
  Physics Letters A} {\bf 153} (1991), no.~6 321 -- 325.

\bibitem{PhysRevD.79.044027}
G.~V. Steeg and N.~C. Menicucci, {\it Entangling power of an expanding
  universe},  {\em Phys. Rev. D} {\bf 79} (Feb, 2009) 044027.

\bibitem{Smith2016topology}
E.~Mart\'{\i}n-Mart\'{\i}nez, A.~R.~H. Smith, and D.~R. Terno, {\it Spacetime
  structure and vacuum entanglement},  {\em Phys. Rev. D} {\bf 93} (Feb, 2016)
  044001.

\bibitem{Ng:2018drz}
K.~K. Ng, R.~B. Mann, and E.~Martín-Martínez, {\it {Unruh-DeWitt detectors
  and entanglement: The anti-de Sitter space}},  {\em Phys. Rev. D} {\bf 98}
  (2018), no.~12 125005, [\href{http://arxiv.org/abs/1809.06878}{{\tt
  arXiv:1809.06878}}].

\bibitem{Henderson:2018lcy}
L.~J. Henderson, R.~A. Hennigar, R.~B. Mann, A.~R. Smith, and J.~Zhang, {\it
  {Entangling detectors in anti-de Sitter space}},  {\em JHEP} {\bf 05} (2019)
  178, [\href{http://arxiv.org/abs/1809.06862}{{\tt arXiv:1809.06862}}].

\bibitem{Henderson:2017yuv}
L.~J. Henderson, R.~A. Hennigar, R.~B. Mann, A.~R.~H. Smith, and J.~Zhang, {\it
  {Harvesting Entanglement from the Black Hole Vacuum}},  {\em Class. Quant.
  Grav.} {\bf 35} (2018), no.~21 21LT02,
  [\href{http://arxiv.org/abs/1712.10018}{{\tt arXiv:1712.10018}}].

\bibitem{Davies:1976hi}
P.~Davies and S.~Fulling, {\it Radiation from a moving mirror in
  two-dimensional space-time conformal anomaly},  {\em Proc. Roy. Soc. Lond. A}
  {\bf A348} (1976) 393--414.

\bibitem{Davies:1977yv}
P.~C.~W. Davies and S.~A. Fulling, {\it {Radiation from Moving Mirrors and from
  Black Holes}},  {\em Proc. Roy. Soc. Lond.} {\bf A356} (1977) 237--257.

\bibitem{fullingpage}
A.~A. Svidzinsky, J.~S. Ben-Benjamin, S.~A. Fulling, and D.~N. Page, {\it
  Excitation of an atom by a uniformly accelerated mirror through virtual
  transitions},  {\em Phys. Rev. Lett.} {\bf 121} (Aug, 2018) 071301.

\bibitem{Good:2013lca}
M.~R.~R. Good, P.~R. Anderson, and C.~R. Evans, {\it Time dependence of
  particle creation from accelerating mirrors},  {\em Phys. Rev. D} {\bf 88}
  (Jul, 2013) 025023.

\bibitem{Good:2016oey}
M.~R.~R. Good, P.~R. Anderson, and C.~R. Evans, {\it {Mirror Reflections of a
  Black Hole}},  {\em Phys. Rev.} {\bf D94} (2016), no.~6 065010,
  [\href{http://arxiv.org/abs/1605.06635}{{\tt arXiv:1605.06635}}].

\bibitem{Good:2016atu}
M.~R.~R. Good, K.~Yelshibekov, and Y.~C. Ong, {\it {On Horizonless Temperature
  with an Accelerating Mirror}},  {\em JHEP} {\bf 03} (2017) 013,
  [\href{http://arxiv.org/abs/1611.00809}{{\tt arXiv:1611.00809}}].

\bibitem{Cong:2018vqx}
W.~Cong, E.~Tjoa, and R.~B. Mann, {\it {Entanglement Harvesting with Moving
  Mirrors}},  {\em JHEP} {\bf 06} (2019) 021,
  [\href{http://arxiv.org/abs/1810.07359}{{\tt arXiv:1810.07359}}]. [Erratum:
  JHEP 07, 051 (2019)].

\bibitem{Johansson:2013asa}
J.~Johansson, G.~Johansson, C.~Wilson, P.~Delsing, and F.~Nori, {\it
  {Nonclassical microwave radiation from the dynamical Casimir effect}},  {\em
  Phys. Rev. A} {\bf 87} (2013), no.~4 043804,
  [\href{http://arxiv.org/abs/1207.1988}{{\tt arXiv:1207.1988}}].

\bibitem{Chen:2017prl}
P.~Chen and G.~Mourou, {\it Accelerating plasma mirrors to investigate black
  hole information loss paradox},  {\em Phys. Rev. Lett.} {\bf 118} (January,
  2017) 045001.

\bibitem{Chen:2020sir}
P.~Chen and G.~Mourou, {\it {Trajectory of a flying plasma mirror traversing a
  target with density gradient}},  \href{http://arxiv.org/abs/2004.10615}{{\tt
  arXiv:2004.10615}}.

\bibitem{Good:2020byh}
M.~R. Good, A.~Zhakenuly, and E.~V. Linder, {\it {The mirror at the edge of the
  universe: Reflections on an accelerated boundary correspondence with de
  Sitter cosmology}},  {\em Phys. Rev. D} (2020)
  [\href{http://arxiv.org/abs/2005.03850}{{\tt arXiv:2005.03850}}].

\bibitem{Good:2020uff}
M.~Good and E.~Abdikamalov, {\it {Radiation from an inertial mirror horizon}},
  {\em Universe} {\bf 6} (2020) 131,
  [\href{http://arxiv.org/abs/2008.08776}{{\tt arXiv:2008.08776}}].

\bibitem{Good:2020fjz}
M.~R. Good, J.~Foo, and E.~V. Linder, {\it {Accelerating boundary analog of a
  Kerr black hole}},  \href{http://arxiv.org/abs/2006.01349}{{\tt
  arXiv:2006.01349}}.

\bibitem{Good:2018aer}
M.~R. Good and E.~V. Linder, {\it {Finite Energy but Infinite Entropy
  Production from Moving Mirrors}},  {\em Phys. Rev. D} {\bf 99} (2019), no.~2
  025009, [\href{http://arxiv.org/abs/1807.08632}{{\tt arXiv:1807.08632}}].

\bibitem{Fabbri}
A.~Fabbri and J.~Navarro-Salas, {\em Modeling Black Hole Evaporation}.
\newblock Imperial College Press, 2005.

\bibitem{Good:2019tnf}
M.~R. Good, E.~V. Linder, and F.~Wilczek, {\it {Moving mirror model for
  quasithermal radiation fields}},  {\em Phys. Rev. D} {\bf 101} (2020), no.~2
  025012, [\href{http://arxiv.org/abs/1909.01129}{{\tt arXiv:1909.01129}}].

\bibitem{Good:2018zmx}
M.~R.~R. Good, {\it {Spacetime Continuity and Quantum Information Loss}},  {\em
  Universe} {\bf 4} (2018), no.~11 122.

\bibitem{Ford:1990id}
L.~Ford, {\it {Constraints on negative energy fluxes}},  {\em Phys. Rev. D}
  {\bf 43} (1991) 3972--3978.

\bibitem{PhysRevD.94.104041}
K.~K. Ng, R.~B. Mann, and E.~Mart\'{\i}n-Mart\'{\i}nez, {\it Equivalence
  principle and qft: Can a particle detector tell if we live inside a hollow
  shell?},  {\em Phys. Rev. D} {\bf 94} (Nov, 2016) 104041.

\bibitem{harvest}
W.~Cong, E.~Tjoa, and R.~B. Mann, {\it Entanglement harvesting with moving
  mirrors},  {\em Journal of High Energy Physics} {\bf 2019} (Jun, 2019) 21.

\bibitem{Smith_2014}
A.~R.~H. Smith and R.~B. Mann, {\it Looking inside a black hole},  {\em
  Classical and Quantum Gravity} {\bf 31} (apr, 2014) 082001.

\bibitem{EoF}
W.~K. Wootters, {\it Entanglement of formation and concurrence},  {\em Quantum
  Info. Comput.} {\bf 1} (Jan., 2001) 27–44.

\bibitem{Cong:2020crf}
W.~Cong, J.~Bicak, D.~Kubiznak, and R.~B. Mann, {\it {Quantum Distinction of
  Inertial Frames: Local vs. Global}},
  \href{http://arxiv.org/abs/2003.09719}{{\tt arXiv:2003.09719}}.

\bibitem{Ng:2016hzn}
K.~K. Ng, R.~B. Mann, and E.~Martin-Martinez, {\it {The equivalence principle
  and QFT: Can a particle detector tell if we live inside a hollow shell?}},
  {\em Phys. Rev.} {\bf D94} (2016), no.~10 104041,
  [\href{http://arxiv.org/abs/1606.06292}{{\tt arXiv:1606.06292}}].

\bibitem{DeWitt:1975ys}
B.~S. DeWitt, {\it {Quantum Field Theory in Curved Space-Time}},  {\em Phys.
  Rept.} {\bf 19} (1975) 295--357.

\bibitem{Good:2018ell}
M.~R. Good, Y.~C. Ong, A.~Myrzakul, and K.~Yelshibekov, {\it {Information
  preservation for null shell collapse: a moving mirror model}},  {\em Gen.
  Rel. Grav.} {\bf 51} (2019), no.~7 92,
  [\href{http://arxiv.org/abs/1801.08020}{{\tt arXiv:1801.08020}}].

\bibitem{Lin:2008jj}
S.-Y. Lin, C.-H. Chou, and B.~Hu, {\it {Disentanglement of two harmonic
  oscillators in relativistic motion}},  {\em Phys. Rev. D} {\bf 78} (2008)
  125025, [\href{http://arxiv.org/abs/0803.3995}{{\tt arXiv:0803.3995}}].

\bibitem{Ostapchuk:2011ud}
D.~C. Ostapchuk, S.-Y. Lin, R.~B. Mann, and B.~Hu, {\it {Entanglement Dynamics
  between Inertial and Non-uniformly Accelerated Detectors}},  {\em JHEP} {\bf
  07} (2012) 072, [\href{http://arxiv.org/abs/1108.3377}{{\tt
  arXiv:1108.3377}}].

\bibitem{Martin-Martinez:2015qwa}
E.~Martin-Martinez, A.~R.~H. Smith, and D.~R. Terno, {\it {Spacetime structure
  and vacuum entanglement}},  {\em Phys. Rev. D} {\bf 93} (2016), no.~4 044001,
  [\href{http://arxiv.org/abs/1507.02688}{{\tt arXiv:1507.02688}}].

\bibitem{Bianchi:2014qua}
E.~Bianchi and M.~Smerlak, {\it {Entanglement entropy and negative energy in
  two dimensions}},  {\em Phys. Rev. D} {\bf 90} (2014), no.~4 041904,
  [\href{http://arxiv.org/abs/1404.0602}{{\tt arXiv:1404.0602}}].

\bibitem{Bianchi:2014bma}
E.~Bianchi, T.~De~Lorenzo, and M.~Smerlak, {\it {Entanglement entropy
  production in gravitational collapse: covariant regularization and solvable
  models}},  {\em JHEP} {\bf 06} (2015) 180,
  [\href{http://arxiv.org/abs/1409.0144}{{\tt arXiv:1409.0144}}].

\bibitem{Good:2017ddq}
M.~R. Good and E.~V. Linder, {\it {Eternal and evanescent black holes and
  accelerating mirror analogs}},  {\em Phys. Rev. D} {\bf 97} (2018), no.~6
  065006, [\href{http://arxiv.org/abs/1711.09922}{{\tt arXiv:1711.09922}}].

\bibitem{Chen:2017lum}
P.~Chen and D.-h. Yeom, {\it {Entropy evolution of moving mirrors and the
  information loss problem}},  {\em Phys. Rev. D} {\bf 96} (2017), no.~2
  025016, [\href{http://arxiv.org/abs/1704.08613}{{\tt arXiv:1704.08613}}].

\bibitem{Good:2020nmz}
M.~R.~R. Good, {\it {Extreme Hawking Radiation}},  {\em Phys. Rev. D} {\bf 101}
  (2020), no.~10 104050, [\href{http://arxiv.org/abs/2003.07016}{{\tt
  arXiv:2003.07016}}].

\bibitem{purity}
F.~Wilczek, {\it {Quantum purity at a small price: Easing a black hole
  paradox}},  in {\em {International Symposium on Black holes, Membranes,
  Wormholes and Superstrings}}, pp.~1--21, 2, 1993.
\newblock \href{http://arxiv.org/abs/hep-th/9302096}{{\tt hep-th/9302096}}.

\end{thebibliography}\endgroup

\end{document}